\documentclass[runningheads,fleqn]{cl2emult}
\usepackage{graphicx}
\usepackage{makeidx}
\usepackage{subeqnar}
\usepackage{multicol}
\usepackage{cropmark}

\begin{document}

\title{Squeezing and feedback}
\author{H.M. Wiseman}
\institute{Department of Physics, University of Queensland, Queensland
4072 Australia \\ and \\
School of Science, Griffith University, Nathan, Queensland 4111 Australia 
\footnote{Permanent address. E-mail: H.Wiseman@gu.edu.au}}
\maketitle

\let\leftlegendglue\relax
\newdimen\mathindent
\mathindent = \leftmargini

\newcommand{\beq}{\begin{equation}}
\newcommand{\eeq}{\end{equation}}
\newcommand{\bqa}{\begin{eqnarray}}
\newcommand{\eqa}{\end{eqnarray}}
\newcommand{\erf}[1]{(\ref{#1})}
\newcommand{\nn}{\nonumber}
\newcommand{\dg}{^\dagger}
\newcommand{\smallfrac}[2]{\mbox{$\frac{#1}{#2}$}}
\newcommand{\bra}[1]{\langle{#1}|}
\newcommand{\ket}[1]{|{#1}\rangle}
\newcommand{\ip}[1]{\langle{#1}\rangle}
\newcommand{\sch}{Schr\"odinger }
\newcommand{\schs}{Schr\"odinger's }
\newcommand{\hei}{Heisenberg }
\newcommand{\heis}{Heisenberg's }
\newcommand{\half}{\smallfrac{1}{2}}
\newcommand{\bl}{{\bigl(}}
\newcommand{\br}{{\bigr)}}
\newcommand{\ito}{It\^o }
\newcommand{\str}{Stratonovich }
\newcommand{\rt}[1]{\sqrt{#1}\,}
\newcommand{\ve}{\varepsilon}

\begin{abstract}
	Electro-optical feedback has many features in common with optical
	nonlinearities and hence is relevant to the generation of squeezing.
	First, I discuss theoretical and experimental results for
    traveling-wave
	feedback, emphasizing how the ``in-loop'' squeezing
	(also known as ``squashing'') differs from free squeezing.
	Although such feedback, based on ordinary (demolition) photodetection
	cannot create free squeezing, it can be used to manipulate it.
	Then I treat feedback based on nonlinear quantum optical measurements
	(of which non-demolition
	measurements are one example). These {\em are} able to produce free
	squeezing, as shown in a number of experiments. Following that
	I discuss theories showing that intracavity squeezing can be
	increased using ordinary feedback, and produced using QND-based
	feedback. Finally, I return to ``squashed'' fields and present
	recent results showing that the reduced in-loop fluctuations can
	suppress atomic decay in a manner
	 analogous to the effect for squeezed fields.
\end{abstract}

\section{Introduction}

In its broadest conception, feedback could be defined to be any
mechanism by which a system acts upon itself, via an intermediate system.
This definition would classify, for example, a nonlinear refractive
index as feedback on a beam of light. The light polarizes
the medium in which it is propagating, which then affects the
propagation of the light in that medium. If one is interested in the light
as a quantum system, then such feedback can be modeled by modifying the
Hamiltonian for the field. This is a well-known mechanism for generating
squeezed states of light \cite{WalMil94}.

In this chapter I am concerned with a different concept of
feedback, in which the intermediate system is external to the system of
interest. That is to say, the system is an {\em open} system, in constant
interaction with its environment. Feedback will occur if the change in
the environment due to the system is significant in affecting the
system's dynamics. Since an environment is by definition large compared
to the system, it is usual for this feedback to be ignored. This is the
essence of the Born, or perturbative, approach to open systems \cite{Gar91}.
However, the system's environment can be deliberately engineered so that
the feedback is important.

One obvious way to achieve this is for the environment to include a
measurement
apparatus which detects the influence of the system on its
surroundings, a
device to amplify this measurement, and a mechanism by which this
amplified
signal controls the dynamics of the system. It is also possible to
engineer a
feedback mechanism which does not involve a measurement device, but
rather some
more direct form of back-coupling from the environment to the system.
In classical mechanics it is a moot point whether a device is
designated
a ``measurement apparatus''. However in quantum mechanics the special
role of
measurement implies that the direct back-coupling may be quite
distinct from feedback via measurement \cite{WisMil94b}.
This is one reason why a peculiarly quantum
theory of feedback is necessary, and interesting.
In this chapter I will be
concerned only with measurement-based feedback.

The history of feedback in quantum optics goes back to the
observation
of sub-shot-noise fluctuations in an in-loop photocurrent in the mid
1980's by two groups \cite{WalJak85a,MacYam85}. The theory of this
phenomenon was soon addressed
by Yamamoto and co-workers \cite{HauYam86,YamImoMac86}, and
Shapiro {\em et al} \cite{Sha87}. The central question they were
addressing was whether this feedback was producing real squeezing, a
question
whose answer is not as straightforward as might be thought.
These treatments were based in the Heisenberg picture and used
quantum Langevin
equations \cite{Gar91} where necessary to describe the evolution of
system operators. They treated
the quantum noise only within a linearized approximation. Although
this
approximation is probably valid for all quantum
optical feedback experiments performed so far, it would not be valid
in the ``deep quantum''
regime involving few photons and non-perturbative couplings, as is
being explored in the so-called ``cavity quantum electrodynamics''
experiments \cite{CarTiaRenAls94}.

More recently an alternative approach to quantum feedback has been
proposed by
myself and Milburn \cite{WisMil93b,WisMil94a}, and developed fully by
myself \cite{Wis94a}. This is based on the theory of quantum
trajectories
\cite{Car93b,GarParZol92,WisMil93c}, which is an application of
quantum
measurement theory to continuously monitored open quantum systems. By
treating
the measurement explicitly, this theory translates the quantum noise
of the bath
into classical noise in the record of detections. It can be shown to
be equivalent to an exact (unlinearized) 
quantum Langevin treatment \cite{Wis94a}.
The advantage of the quantum trajectory method is that it allows
arbitrary feedback
to be treated by the theory, at least numerically. A particular limit
of
interest is that of Markovian feedback, in which the feedback
dynamics can be
modeled using a master equation. This  result was not obtained by
the authors using the quantum Langevin treatment.

A third approach \cite{Pli94} to feedback in quantum optics is to use
the Glauber-Sudarshan $P$ function \cite{Gla63a,Gla63b,Sud63}, a
quasi-probability distribution. In this theory, the fields are
given an essentially classical description, but
negative probabilities are allowed in order to take into account
quantum
correlations \cite{Pli94}. This theory is just as easy to use as the
quantum
Langevin or quantum trajectory theories  when the system
dynamics which can be linearized. However, like the quantum Langevin
approach, it is usually intractable when the linearization
approximation cannot be made. I will not discuss this theory further
in this chapter.

A fourth approach is to treat the electromagnetic field as a stream
of point-like particles (photons) traveling at the speed of light.
This is essentially a classical approach, which cannot describe phase
properties of the fields, but which is adequate if one is interested
only in intensity statistics. Formally, the in-loop photon arrivals
become a self-excited classical point process. This theory was used
by Shapiro {\em et al.}\cite{Sha87} in addition to their quantum
operator theory. Similar ideas have subsequently been used by other
authors \cite{Tro91,HeiMer93}. Like the other three
approaches mentioned above this approach is easily
applicable to linearized systems, but unlike them
it does not give a full quantum description of the
in-loop field. Again, I will not discuss this
theory further in this chapter.

In the remainder of
this chapter I have alternated `theory' sections, which introduce
the mathematical apparatus necessary for describing quantum feedback,
with `application' sections, which use the
theory to investigate
squeezing, and, where appropriate, discuss experimental results.
In both of these parallel streams the
material is presented in roughly the order in which it was
developed, but the two streams are not synchronous. 

First I introduce
continuum fields, and then show how linearization allows feedback
onto those fields to be treated analytically, yielding noise spectra
for in-loop and out-of-loop measurements. Next I discuss the
interaction of continuum fields with a localized quantum system,
giving rise to quantum Langevin equations for system operators. This
theory is used to describe nonlinear measurements (such as QND
measurements) of continuum fields,
and feedback based on the results of these measurements. An
alternative to the quantum Langevin description is one based on
quantum
trajectories. This is most useful for illuminating feedback onto the
localized systems, and I use it to investigate intracavity squeezing.
In the Markovian limit the quantum trajectory picture of feedback
allows one to derive a feedback master equation. This is of most use
for describing dynamics which cannot be linearized, such as that of a
strongly driven two-level atom. This turns out to be precisely what
is needed to revisit the question of in-loop squeezing in terms of
what the atom `sees'.

\section{Continuum Fields}

\subsection{Canonical Quantization}
\label{sec:cq}

Let the fundamental
field be the vector potential
${\bf A}({\bf r},t)$ in the Coulomb gauge
\beq \label{cg}
\nabla \cdot {\bf A}({\bf r},t) = 0.
\eeq
The free Lagrangian density for this
field is \cite{CohDupGry89}
\begin{equation}
	$\pounds$ = \frac{\varepsilon_0}{2} \left( {\bf E}^2 -
	c^2 {\bf B}^2 \right),
	\label{341lagden}
\end{equation}
where $\varepsilon_0$ is the permittivity of free space and $c$ is
the speed of light. The electric ${\bf E}$ and magnetic ${\bf B}$
fields are defined by
\begin{equation}
	{\bf E}= - \dot{\bf A} \; ; \;\; {\bf B} =
	\nabla \times {\bf A}.
\end{equation}
From \erf{341lagden}, the canonical field to ${\bf A}$ is
$-\varepsilon_0 {\bf E}$.
Thus, in quantizing the field, these obey the canonical
commutation relations
\begin{equation}
	[A_j({\bf r},t),E_k({\bf r}',t)] = -
	{i}\frac{\hbar}{\varepsilon_0}
	 \delta_{jk}\delta^3_{\perp} ({\bf r}-{\bf r}') .
	\label{341ccr}
\end{equation}
Here $\delta^3_{\perp}$ denotes a three-dimensional {\em transverse}
Dirac delta-function, which is necessary to be compatible with the
constraint of \erf{cg} \cite{CohDupGry89}. Note that the Heisenberg
picture operators in the canonical commutation relations are at equal
times. In the Schr\"odinger picture, the same relations hold, but the
time argument is omitted. The Euler-Lagrange (which is also the
Heisenberg) equation of motion from \erf{341lagden} is the wave
equation
\begin{equation}
	\ddot{\bf A}=c^2 \nabla^2 {\bf A}.
\end{equation}

Now consider the case of a beam of polarized light. That is to say,
consider only one component $A$ of ${\bf A}$ and let its spatial
variation be
confined to one direction, say $z$. This simplifies the analysis, and
is
also appropriate for determining the inputs and outputs of a quantum
optical cavity. In reality, the transverse spatial extent of the beam
would be confined to some area $\Lambda$ which is determined by the
area of the
optical components involved \cite{Gar91}. However, as long as the $x$
and
$y$ extensions are much greater than a wavelength, the beam can be
approximated by plane waves. The appropriate wave equation is
\begin{equation}
	\ddot{A}=c^2 \partial_z^2 {A},
	\label{341waveq2}
\end{equation}
of which I am interested only in the forward propagating solutions
\begin{equation}
	A(z,t+t')=A(z-ct',t).
	\label{341wavesoln1}
\end{equation}
If the field is reflected off a cavity mirror (say at $z=0$) then the
direction
of $z$ will change at the point of reflection. This is why only one
direction
of propagation need be considered. The field for $z<0$ is incoming
and
that for $z>0$ is outgoing. The canonical commutation relation is now
\begin{equation}
	[A(z,t),E(z',t)]=-{i}\frac{\hbar}{\varepsilon_0
\Lambda}\delta(z-z').
	\label{341ccr2}
\end{equation}

Solutions for $A$ and $E$ satisfying the wave equation
(\ref{341waveq2})
can be constructed using the annihilation and creation operators for
the modes of frequency $\omega$, which satisfy
\begin{equation}
	[a(\omega),a\dg(\omega')]=\delta(\omega-\omega').
\end{equation}
They are
\begin{eqnarray}
A(z) & = & \sqrt{\frac{\hbar}{\varepsilon_0 \Lambda 2\pi c}}\int_0^\infty
d\omega \frac{1}{\sqrt{2\omega}} \left\{ a(\omega) \exp[-{i}
\omega(t-z/c)] + {\rm H.c.}\right\}, \label{341A} \\
E(z) & = & \sqrt{\frac{\hbar}{\varepsilon_0 \Lambda 2\pi c}}\int_0^\infty
d\omega \sqrt{\frac{\omega}{2}} \left\{ {i}a(\omega) \exp[-{\rm i}
\omega(t-z/c)] + {\rm H.c.}\right\}. \label{341E}
\end{eqnarray}

This expression for the fields in terms of annihilation and creation
operators for a continuum of modes defines the sense in which they
are composed of photons  of definite frequency. However, this
sense is quite unlike the naive picture of a beam of light made
up of (possibly different frequencies of) photons, hurtling through
space at the speed of light. Each mode is spread over all space, so there
is no way in which a photon, as an excitation of such a mode, can move
at all. To define an annihilation operator $b(z,t)$ for a localized
photon of a particular frequency, it would be
necessary to sum many different mode operators. Such operators can be
defined, with slight variations in the details of the definition
\cite{GarParZol92,GarCol85}. The various definitions
are effectively equivalent in application to quantum optical
problems.
The authors of \cite{GarParZol92,GarCol85} construct the
localized annihilation operator
from the mode annihilation operators $a(\omega)$. Here, just for
variation, I am
introducing a different definition for $b(z,t)$, constructed from the
original fields in space-time, $A(z,t)$ and $E(z,t)$.

As established above, $A$ and $-\varepsilon_{0}E$ are canonically
conjugate variables at each
point in space-time.  Motivated by the analogy with position and
momentum,
a local annihilation operator for an oscillator of angular frequency
$\omega_0$ can be defined as
\begin{equation}
	b(z,t) = \exp[{i}\omega_0(t-z/c)]
\sqrt{\frac{\Lambda c}
	{\hbar}} 	\left[ \sqrt{\frac{\omega_0\varepsilon_0}{2}}A(z,t)
	-\frac{i\varepsilon_{0}}{\sqrt{2\omega_0\varepsilon_0}}E(z,t)
\right].
	\label{341bzt1}
\end{equation}
In terms of the mode operators, $b(z,t)$ is given by
\begin{eqnarray}
b(z,t)&=& \frac{1}{\sqrt{2\pi}} \int_0^\infty d\omega
\left\{\frac{\omega_0+\omega}{2\sqrt{\omega_0\omega}} a(\omega)
\exp[{i}(\omega_0-\omega)(t-z/c)]\right. \nonumber \\
&&+ \left. \frac{\omega_0-\omega}{2\sqrt{\omega_0\omega}} a\dg(\omega)
\exp[{i}(\omega_0+\omega)(t-z/c)]\right\}. \label{341bzt2}
\end{eqnarray}
If the beam contains only photons of a frequency near $\omega_{0}$
then it is apparent from \erf{341bzt2} that we can approximate
$b(z,t)$ by
\beq
b(z,t) \approx \frac{1}{\sqrt{2\pi}} \int_0^\infty d\omega\,
	 a(\omega) \exp[{i}(\omega_0-\omega)(t-z/c)].
\eeq
From this we can calculate
\begin{eqnarray}
	[b(z,t),b\dg(z',t)] &\approx&
	\frac{1}{2\pi} \int_{0}^{\infty}d\omega
	\exp[{i}(\omega_0-\omega)(z'-z)/c] \\
	&\approx& c\delta(z-z'),
	\label{341crb}
\end{eqnarray}
which explains the choice of normalization in \erf{341bzt1}.
Note that the approximations involved here apply only if all of the
light
is at a frequency close to $\omega_{0}$. Thus, the width of the Dirac
$\delta$ function in \erf{341crb} should be understood to be much
greater than a wavelength of light $2\pi c / \omega_{0}$.

The rotating exponential $\exp[{i}\omega_0(t-z/c)]$ means that
 a ``localized photon'' of frequency
$\omega_0$ has a slowly varying annihilation operator  $b(z,t)$.
This operator also has the same property as the vector potential
(\ref{341wavesoln1}), obeying
\begin{equation}
	b(z,t+t')=b(z-ct',t)
	\label{wavesoln2}
\end{equation}
in free space.
 If only frequencies near $\omega_0$ are
significantly excited, then the time-flux of energy can be easily
seen to
be
\begin{equation}
	W(z,t) \simeq \hbar \omega_0 b\dg(z,t) b(z,t).
\end{equation}
Thus, the annihilation operator $b(z,t)$ conforms to one's naive
expectations, with $b\dg(z,t) b(z,t)$ being the photon flux (photons 
per unit time) passing $z$ at time $t$.

\subsection{Photodetection}

From the above discussion it should be apparent that it is not
sensible to talk about a photodetector for photons of frequency
$\omega_{0}$ which has a response time comparable to or smaller 
 than $\omega_{0}^{-1}$.
Therefore for practical purposes a photodetector is equivalent to an
energy flux meter. In either case, as long as we are not interested
in times comparable to $\omega_{0}^{-1}$, we can assume that the
signal
produced by an ideal photodetector at position $z_{1}$ is given by
the operator
\beq \label{intensity}
I(t) = b\dg_{1}(t)b_{1}(t),
\eeq
where $b_{1}(t) \equiv b(z_{1},t)$. Here I have ignored any factors
of electric charge {\em etc.} which are sometimes included but which are
actually nominal.

In experiments involving lasers, it is often the case (or
at least it is harmless to assume \cite{Mol96}) that $b_{1}(t)$ has
a mean amplitude $\beta = \ip{b_{1}(t)}$. Without loss of generality,
I will take $\beta$ to be real. In all that follows I will also assume 
that we are considering stationary statistics. That is, we are taking 
the long time limit of a system with a stationary state. 
In that case, {\em only if} the correlations
of interest in the intensity of the beam of
light  have a characteristic time satisfying
\beq
\tau_{\rm cor} \gg |\beta|^{-2},
\eeq
 is it permissible to linearize \erf{intensity}. This means
approximating it by
\beq \label{defdI}
I(t) = \beta^{2} + \delta I(t) = \beta^{2}+\beta X_{1}(t),
\eeq
where
\beq
X_{1}(t) = b_{1}(t) + b_{1}\dg(t) - 2\beta
\eeq
is the amplitude quadrature fluctuation operator for the continuum
field. For the linearization to be valid the
fluctuations must be small as well as slow:
\beq
\ip{X_{1}(t+\tau)X_{1}(t)} \ll \beta^{2} 
\textrm{ for }\tau \sim \tau_{\rm corr}.
\eeq

It is useful also to define the phase quadrature fluctuation operator
\beq
Y_{1}(t) = -ib_{1}(t) + ib_{1}\dg(t).
\eeq
For free fields, which obey \erf{wavesoln2},
these obey the commutation relations
\beq
[X_{1}(t),Y_{1}(t')] = 2i\delta(t-t'). \label{comr1}
\eeq
If we define the Fourier transformed operator
\beq
\tilde{X}_{1}(\omega) = \int_{-\infty}^{\infty} dt
X_{1}(t) e^{-i\omega t}
\eeq
and similarly for $\tilde{Y}_{1}(\omega)$ then
\beq
[\tilde{X}_{1}(\omega),\tilde{Y}_{1}(\omega')]
= 4\pi i \delta (\omega + \omega'). \label{comr2}
\eeq
For stationary statistics as we are considering, 
$\ip{X_{1}(t)X_{1}(t')}$ is a function of $t-t'$ only. From this it 
follows that 
\beq \label{almosp}
\ip{\tilde{X}_{1}(\omega)\tilde{X}_{1}(\omega')} \propto 
\delta(\omega+\omega').
\eeq

Because of the singularities in equations~(\ref{comr2}) and 
(\ref{almosp}), to obtain a finite uncertainty relation it 
is more useful to consider
the spectrum
\bqa
S_{1}^{X}(\omega) &=& \frac{1}{2\pi}
\int_{-\infty}^{\infty} \ip{\tilde{X}_{1}(\omega )
\tilde{X}_{1}(-\omega')} d\omega' \label{defspec} \\
&=& \int_{-\infty}^{\infty} e^{-i\omega t} \ip{ X(t)X(0)} dt \,=\,
\ip{\tilde{X}_{1}(\omega)X_{1}(0)}.
\eqa
Then it can be shown that for a stationary free field \cite{Sha87},
\beq
S_{1}^{X}(\omega) S_{1}^{Y}(\omega) \geq 1. \label{ur3}
\eeq
From this it is obvious that a coherent continuum field
\cite{Gla63a,Gla63b,Sud63}  is one such that for all $\omega$
\beq
S_{1}^{Q}(\omega)=1,
\eeq
where $Q=X$ or $Y$ (or any intermediate quadrature). This is known as
the {\em standard quantum limit} or {\em shot-noise limit}. A {\em
squeezed}
continuum field is one such that, for some $\omega$ and some $Q$,
\beq
S_{1}^{Q}(\omega) < 1.
\eeq

The physical significance of $S_{1}^{X}(\omega)$ is apparent from
\erf{defdI}: it can be experimentally determined as
\beq \label{exptdet}
S_{1}^{X}(\omega) = \ip{I(t)}^{-1}
\int_{-\infty}^{\infty} e^{-i\omega t} \ip{ I(t),I(0)} dt,
\eeq
where $\ip{A,B}=\ip{AB} - \ip{A}\ip{B}$. In fact it is possible to
determine $S_{1}^{Q}(\omega)$ for any quadrature $Q$ in a similar
way. Putting the field of interest through a low-reflectivity beam
splitter, while reflecting another field with a large
coherent amplitude off the same beam splitter, a coherent amplitude
can be added to the beam of interest. If this contribution is
sufficiently large it
will dominate the total coherent amplitude of the beam. Since the
added component can have any chosen phase with respect to the
original beam, the new linearized intensity fluctuation operator will
be proportional to  any chosen quadrature fluctuation operator. This
technique is known as homodyne detection. In practice, balanced
homodyne detection using a 50--50 beam splitter and two
photodetectors
is preferable, but the principle is the same \cite{YueSha80}.
\label{secII}

\section{In-loop ``Squeezing''}
\label{secils}

\subsection{Description of the Device}
\label{secdd}

\begin{figure}
\includegraphics[width=1\textwidth]{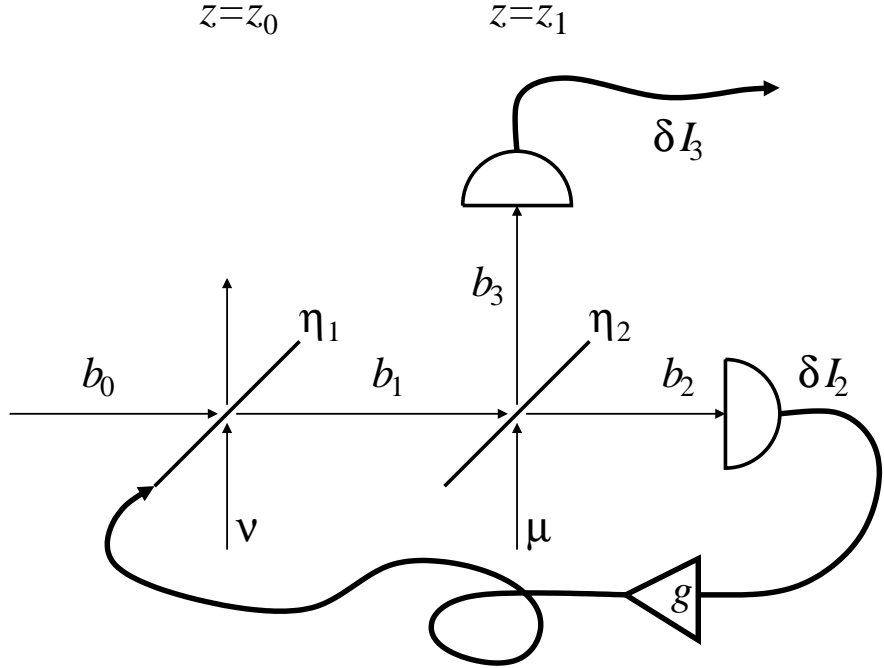}
\caption{Diagram for a traveling-wave feedback experiment. Traveling
fields are denoted $b$ and photocurrent fluctuations $\delta I$.
The first beam splitter
transmittance $\eta_{1}$ is variable, the second $\eta_{2}$ fixed.
The two vacuum field inputs are
denoted $\nu$ and $\mu$}
\label{wfig1}
\end{figure}

The simplest form of quantum optical feedback is shown in 
Fig.~\ref{wfig1}.
This was the scheme considered by Shapiro {\em et al.}. In our
notation, we begin with a field $b_{0}=b(z_{0},t)$ as shown in the
diagram.
We will take this field to have stationary statistics with
mean amplitude $\beta$ and
fluctuations
\beq
\half[ X_{0}(t) + iY_{0}(t)] = b_{0}(t) - \beta
\eeq
characterized by arbitrary spectra
$S_{0}^{X}(\omega),S_{0}^{Y}(\omega)$. This
field is then passed through a beam splitter of transmittance
$\eta_{1}(t)$. By unitarity, the diminution in the transmitted field
by a
factor $\sqrt{\eta_{1}(t)}$ must be accompanied by the addition of
vacuum
noise from the other port of the beam splitter \cite{WalMil94}.
The transmitted field is
\beq \label{etat}
b_{1}(t) = \sqrt{\eta_{1}(t-\tau_{1})}\,b_{0}(t-\tau_{1}) +
\sqrt{\bar{\eta}_{1}(t-\tau_{1})}\,\nu(t-\tau_{1}).
\eeq
Here $\tau_{1} = (z_{1} - z_{0})/c$ and I am using the notation
\beq
\bar{\eta} \equiv 1-\eta.
\eeq
The operator $\nu(t)$ represents the vacuum
fluctuations. The vacuum is special case of a coherent
continuum field of vanishing amplitude
$\ip{\nu(t)}=0$, and so is completely characterized by its spectrum
\beq
S_{\nu}^{Q}(\omega) = 1.
\eeq
Since the vacuum fluctuations
are uncorrelated with any other field, and have stationary
statistics, the phase and time arguments for $\nu(t)$ are arbitrary.

The beam-splitter transmittance $\eta_{1}(t)$ in \erf{etat} is
time-dependent. This time-dependence can be achieved experimentally
by a number of means. For example, if the incoming beam is
elliptically
polarized then
an electro-optic modulator (a device with a refractive index
controlled by a current) will alter the orientation of the ellipse. A
polarization-sensitive beam splitter will then control the amount of
the light which is transmitted, as done, for example, in
\cite{Tau95}. As the reader will no doubt have
anticipated, the current used to control the electro-optic modulator
can be derived from a later detection of the light beam, giving rise
to feedback. Writing $\eta_{1}(t)=\bar{\eta}_{1} + \delta\eta_{1}(t)$, 
and assuming that the modulation of the transmittance is
small ($\delta\eta_{1}(t) \ll \eta_{1},\bar{\eta}_{1}$), one can write
\beq \label{lineta}
\sqrt{\eta_{1}(t)} = \sqrt{\eta_{1}} +
(1/\sqrt{\eta_{1}})\half \delta \eta_{1}(t).
\eeq

Continuing to follow the path of the beam in Fig.~\ref{wfig1}, it now enters a
second beam-splitter of constant transmittance $\eta_{2}$. The
transmitted beam annihilation operator
\beq
b_{2}(t) = \sqrt{\eta_{2}}\,b_{1}(t-\tau_{2}) +
\sqrt{\bar{\eta}_{2}}\,\mu(t-\tau_{2}),
\eeq
where $\tau_{2}=(z_{2}-z_{1})/c$ and $\mu(t)$ represents vacuum
fluctuations like $\nu(t)$. The reflected beam operator is
\beq
b_{3}(t) = \sqrt{\bar{\eta}_{2}}\,b_{1}(t-\tau_{2}) -
\sqrt{\eta_{2}}\,\mu(t-\tau_{2}).
\eeq
Using the approximation (\ref{lineta}), the linearized quadrature
fluctuation operators for $b_{2}$  are
\bqa
X_{2}(t) &=& \sqrt{\eta_{2}\eta_{1}}\,X_{0}(t-T_{2})
+ \sqrt{\eta_{2}/\eta_{1}}\,\beta\delta\eta_{1}(t-T_{2}) \nonumber \\
&&+\, \sqrt{\eta_{2}\bar{\eta}_{1}} \,X_{\nu}(t-T_{2})
+ \sqrt{\bar{\eta}_{2}}\,X_{\mu}(t-T_{2}), \label{X2a} \\
Y_{2}(t) &=& \sqrt{\eta_{2}\eta_{1}}\,Y_{0}(t-T_{2})
+ \sqrt{\eta_{2}\bar{\eta}_{1}} \,Y_{\nu}(t-T_{2}) 
+\sqrt{\bar{\eta}_{2}}\,Y_{\mu}(t-T_{2}),
\eqa
where $T_{2} = \tau_{2} + \tau_{1}$. Similarly for $b_{3}$ we have
\bqa
X_{3}(t) &=& \sqrt{\bar{\eta}_{2}\eta_{1}}\,X_{0}(t-T_{2})
+ \sqrt{\bar{\eta}_{2}/\eta_{1}}\,\beta\delta\eta_{1}(t-T_{2})
\nonumber \\
&&+\, \sqrt{\bar{\eta}_{2}\bar{\eta}_{1}} \,X_{\nu}(t-T_{2})
- \sqrt{\eta_{2}}\,X_{\mu}(t-T_{2}), \label{X3a}\\
Y_{3}(t) &=& \sqrt{\bar{\eta}_{2}\eta_{1}}\,Y_{0}(t-T_{2})
+ \sqrt{\bar{\eta}_{2}\bar{\eta}_{1}} \,Y_{\nu}(t-T_{2})
- \sqrt{\eta_{2}}\,Y_{\mu}(t-T_{2}).
\eqa

The mean fields for $b_{2}$ and $b_{3}$ are
$\sqrt{\eta_{1}\eta_{2}}\,
\beta$ and $\sqrt{\eta_{1}\bar{\eta}_{2}}\,\beta$ respectively. Thus,
if
these fields are incident upon photodetectors, the respective
linearized photocurrent
fluctuations are, as explained in Sec.~\ref{secII},
\bqa
\delta I_{2}(t) &=& \sqrt{\eta_{1}\eta_{2}}\,\beta \, X_{2}(t), \\
\delta I_{3}(t) &=& \sqrt{\eta_{1}\bar{\eta}_{2}}\,\beta \, X_{3}(t).
\eqa
Here I have assumed perfect efficiency detectors. To model
inefficient detectors it is necessary to add further beam splitters,
with transmittance equal to the detection efficiency, in front of the
detectors. The effect of this has been considered in detail
in~\cite{Tau95,Mas94}.

Having obtained an expression for $\delta I_{2}(t)$ we are now in a
position to follow the next stage in Fig.~\ref{wfig1} and complete the feedback
loop. We set the modulation in the transmittance of the first
beam-splitter
to be
\beq \label{etatb}
\delta \eta_{1}(t) = \frac{{g}}{\eta_{2}\beta^{2}}
\int_{0}^{\infty} h(t') \delta I_{2}(t-\tau_{0} - t') dt',
\eeq
where ${g}$ is a dimensionless parameter representing the
low-frequency gain of the feedback loop. The response of the
feedback loop, including the electro-optic elements,
is assumed to be linear for small fluctuations and is characterized
by the electronic delay time $\tau_{0}$ and the response function $h(t')$,
which satisfies $h(t)=0$ for $t<0$, $h(t)\geq 0$ for $t>0$
 and $\int_{0}^{\infty} h(t') dt' = 1$.

\subsection{Stability} \label{sec:stab}

Clearly the feedback can only affect the amplitude quadrature $X$.
Putting
\erf{etatb} into \erf{X2a} yields
\bqa
X_{2}(t) &=& \sqrt{\eta_{2}\eta_{1}}\,X_{0}(t-T_{2})
+ {g} \int_{0}^{\infty} h(t')  X_{2}(t-T - t') dt' \nonumber \\
&& +\,\sqrt{\eta_{2}\bar{\eta}_{1}} \,X_{\nu}(t-T_{2})
+ \sqrt{\bar{\eta}_{2}}\,X_{\mu}(t-T_{2}), \label{X2c}
\eqa
where $T = \tau_{0}+ T_{2}= \tau_{0}+\tau_{1}+\tau_{2}$.
This is easy to solve in Fourier space, providing that $X_{2}$ is a
stationary stochastic process. This will only be the case if
the feedback is stable.
Using standard feedback and control theory \cite{Ste90}, the
Nyquist stability criterion is
\beq \label{nyq}
{\rm Re}[s] < 0,
\eeq
where $s$ is any solution of the characteristic equation
\beq \label{cheq}
1-{g} H(s) \exp(-s T) = 0,
\eeq
where $H(s)$ denotes the Laplace transform
$H(s) = \int_{0}^{\infty} dt e^{-st}h(t)$.

First I show that a {\em sufficient} condition for stability is $|g|<1$.
Looking for instability, assume that ${\rm Re}[s]>0$. Then
\beq
\left| H(s)e^{-sT} \right|=
\left|\int_{0}^{\infty} dt e^{-s(t+T)} h(t) \right| \leq
\int_{0}^{\infty} dt  h(t) = 1.
\eeq
Thus under this assumption the characteristic equation
cannot be satisfied for $|{g} | < 1$, so this regime will always be
stable. If ${g} > 1$ then it is not difficult to show that there is
a positive ${s}$ which will solve
\erf{cheq}. Thus it is a {\em necessary} condition to have ${g} < 1$.
If ${g} < -1$, the stability of the feedback depends on $T$ and the
shape of $h(t)$. However, it turns out that it is possible to have
arbitrarily large negative low-frequency feedback (that is, ${g}
\to -\infty$), for any feedback loop delay $T$, provided that $h(t)$
is broad enough. The price to be paid for strong low-frequency
negative feedback is a reduction in the bandwidth of the feedback, the 
width of $|\tilde{h}(\omega)|^{2}$.

To see this, consider the simplest smoothing function $h(t)=\gamma
e^{-\gamma t}$. The condition for marginal stability is that there is
a solution to \erf{cheq} for $s=i\omega$. That is,
\beq \label{rhs}
1={g} \exp({-i\omega T}) \frac{\gamma}{\gamma + i\omega}.
\eeq
For the imaginary part of the right-hand side to vanish we require
\beq \label{te}
\tan \omega T = -\omega/\gamma.
\eeq
As we will see, for large $|{g}|$
we will require $\gamma \ll T^{-1}$ in which case
the solutions on \erf{te} can be approximated by
$\omega_{n}=(2n+1)\pi/2T$. Under the same approximation we can
ignore $\gamma$ compared to $\omega$ in \erf{rhs} to get
\beq
1 = -|{g}| (-i)^{2n+1}\frac{2\gamma T}{i(2n+1)\pi}.
\eeq
Clearly for $n$ odd this cannot be satisfied and so the system will
be stable. However for $n$ even we have
\beq
1= |{g}| \frac{2\gamma T}{(2n+1)\pi},
\eeq
which can be satisfied (indicating marginal stability). In order to
avoid this for all $n$ we require
\beq
\gamma < \frac{\pi}{2 T |{g}|} \ll \frac{1}{T},
\eeq
where here we see that $\gamma \ll T^{-1}$ for large negative ${g}$.
Now the bandwidth of the feedback is
$B  \simeq 2\gamma$. Thus we have finally the
approximate inequality
\beq
B \leq \frac{\pi}{T |{g}|},
\eeq
which shows how a finite delay time $T$ and
large negative feedback $-{g} \gg 1 $ reduces the possible bandwidth of the
feedback.

\subsection{In-loop and Out-of-loop Spectra}
\label{ilols}

Assuming then that the feedback is stable we can solve \erf{X2c} for $X_{2}$ in
the Fourier domain:
\beq \label{X2fou}
\tilde{X}_{2}(\omega) = \exp({-i\omega T_{2}})\frac{
\sqrt{\eta_{2}\eta_{1}}\,\tilde{X}_{0}(\omega) +
\sqrt{\eta_{2}\bar{\eta}_{1}} \,\tilde{X}_{\nu}(\omega) +
\sqrt{\bar{\eta}_{2}}\,\tilde{X}_{\mu}(\omega)
}{1-{g} \tilde{h}(\omega) \exp(-i\omega T)}.
\eeq
From this the amplitude quadrature spectrum is easily found from
\erf{almosp} and \erf{defspec} to be
\bqa
S_{2}^{X}(\omega) &=&
\frac{\eta_{1}\eta_{2}S_{0}^{X}(\omega)+
\eta_{2}\bar{\eta}_{1}S_{\nu}^{X}(\omega)+
\bar{\eta}_{2}S_{\mu}^{X}(\omega)}
{|1-{g} \tilde{h}(\omega) \exp(-i\omega T)|^{2}} \nn \\
&=& \frac{1 + \eta_{1}\eta_{2}[S_{0}^{X}(\omega)-1]}
{|1-{g} \tilde{h}(\omega) \exp(-i\omega T)|^{2}} .\label{S2X}
\eqa

From these formulae the effect of feedback is obvious: it
multiplies the amplitude quadrature spectrum at a given frequency by
the factor $|1-{g} \tilde{h}(\omega) \exp(-i\omega T)|^{-2}$.
At low frequencies, this factor is simply $(1-{g})^{-2}$, which
is why the feedback was classified on this basis into positive
(${g} > 0$) and negative (${g} < 0$) feedback. The former
will increase the noise at low frequency and the latter will decrease
it. However at higher frequencies, and in particular at multiples of
$\pi/T$, the sign of the feedback will reverse and ${g}<0$ will
result in an increase in noise and vice-versa. This is shown clearly
in the theoretical investigations of Shapiro {\em et al.}. All of
these results make perfect sense in the context of classical light
signals, except that in that case we would not worry about vacuum
noise. This is equivalent to assuming that the original noise is far
above the shot-noise limit, so that one can replace
$1 + \eta_{1}\eta_{2}[S_{0}^{X}(\omega)-1]$ by
$\eta_{1}\eta_{2}S_{0}^{X}(\omega)$. This gives the result expected
from classical signal processing: the signal is attenuated by the
beam splitters and either amplified or suppressed by the feedback.

The most dramatic effect is of course for large negative
feedback. For sufficiently large $-{g}$ it is clear that one can
make
\beq
S_{2}^{X}(\omega) < 1
\eeq
for some $\omega$. This effect has been observed experimentally many
times with different systems involving feedback
\cite{WalJak85a,MacYam85,YamImoMac86,Fon91,Tau95,Mas94,You94}. Without a
feedback loop this sub-shot-noise photocurrent
would be seen as evidence for squeezing. However,
there are a number of reasons to be very cautious about applying the
word  squeezing to this  phenomenon. Two of these reasons are
theoretical, and are discussed in the following two sub-sections.
The more practical reason relates to the out-of-loop beam $b_{3}$, which
I will now discuss.

From \erf{X3a}, the $X$ quadrature of the beam $b_{3}$ is, in the
Fourier domain,
\bqa
\tilde{X}_{3}(\omega) &=& \exp({-i\omega T_{2}}) \left[
\sqrt{\bar{\eta}_{2}\eta_{1}}\,\tilde{X}_{0}(\omega)
+ \sqrt{\bar{\eta}_{2}\bar{\eta}_{1}} \,\tilde{X}_{\nu}(\omega)
- \sqrt{\eta_{2}}\,\tilde{X}_{\mu}(\omega) \right] \nonumber \\
&& +\,\sqrt{\bar{\eta}_{2}/\eta_{2}} {g} \tilde{h}(\omega)
\exp(-i\omega T) \tilde{X}_{2}(\omega).
\eqa
Here I have substituted for $\delta\eta_{1}$ in terms of $X_{2}$. Now
using the above expression (\ref{X2fou}) gives
\bqa
\tilde{X}_{3}(\omega) &=& \exp({-i\omega T_{2}})\left\{
\frac{\sqrt{\bar{\eta}_{2}\eta_{1}}\,\tilde{X}_{0}(\omega)+
\sqrt{\bar{\eta}_{2}\bar{\eta}_{1}}\,\tilde{X}_{\nu}(\omega)}
{1-{g} \tilde{h}(\omega) \exp(-i\omega T)}\right.
\nn \\
&& -\,\left.
\frac{\sqrt{\eta_{2}}[1-{g}\tilde{h}(\omega) \exp(-i\omega T)/\eta_{2}]
\tilde{X}_{\mu}(\omega)}{1-{g} \tilde{h}(\omega) \exp(-i\omega
T)}\right\}.
\eqa
This yields the spectrum
\bqa
S_{3}^{X}(\omega) \label{S3a}
&=& \frac{1 + \bar{\eta}_{2}{\eta}_{1}(S_{0}^{X}-1)}{|1-{g} \tilde{h}(\omega)
\exp(-i\omega T)|^{2}} \nonumber \\
& &- \frac{2{\rm Re}[{g}\tilde{h}(\omega) \exp(-i\omega T) ] +
{g}^{2}|\tilde{h}(\omega)|^{2}/\eta_{2}}
{|1-{g} \tilde{h}(\omega) \exp(-i\omega T)|^{2}}.
\eqa
The denominator is identical to that in the in-loop case, as are the
first two terms in the numerator. But there are additional terms in
the numerator which indicate that there is extra noise in the
out-of-loop signal.

The expression (\ref{S3a}) can be rewritten as
\beq
S_{3}^{X}(\omega) = 1 + \frac{\bar{\eta}_{2}{\eta}_{1}(S_{0}^{X}-1)
+ {g}^{2}|\tilde{h}(\omega)|^{2}\bar{\eta}_{2}/\eta_{2}}
{|1-{g} \tilde{h}(\omega) \exp(-i\omega T)|^{2}}.
\eeq
From this it is apparent that, unless the initial beam is amplitude
squeezed (that is, unless $S_{0}^{X}(\omega)<1$ for some $\omega$)
the out-of-loop spectrum will always be greater than the
shot-noise-limit of unity. In other words, it is not possible to
extract the apparent squeezing in the feedback loop by using a beam
splitter. In fact, in the limit of large negative feedback (which
gives the greatest noise reduction in the in-loop signal), the
out-of-loop amplitude spectrum approaches a constant. Considering a
frequency
$\omega$ such that $\tilde{h}(\omega) \exp(-i\omega T)$ is real and
positive, one finds that
\beq \label{limlam}
\lim_{{g}
\to - \infty} S_{3}^{X}(\omega) = \eta_{2}^{-1}.
\eeq
Thus the more light one attempts to extract from the feedback loop,
the higher above shot-noise the spectrum becomes.

This result is counter to an intuition based on classical light
signals, where the effect of a beam splitter is simply to split a
beam so that both outputs would have the same statistics. The reason
this intuition fails is precisely because this is not all that a beam
splitter does; it also introduces vacuum noise which is {\em
anticorrelated} at the two output ports.  The
detector for beam $b_{2}$ measures the amplitude fluctuations
$X_{2}$, which
are a combination of the initial fluctuations $X_{0}$, and the two
vacuum fluctuations $X_{\nu}$ and $X_{\mu}$. The first two of these
are common to the beam $b_{3}$, but the last, $X_{\mu}$, appears with
opposite sign in $X_{3}$. As the negative feedback is turned up, the
first two components are successfully suppressed, but the last is
actually amplified. Note that the result in \erf{limlam} holds no
matter how large $S_{0}(\omega)$ is compared to unity.

\label{secvf}

\subsection{Commutation Relations}

Under normal circumstances (without a feedback loop) one would expect
a sub-shot noise amplitude
spectrum to imply a super-shot-noise phase spectrum. However that is
not what is found from the theory presented here. Rather, the in-loop
phase
quadrature spectrum is unaffected by the feedback, being equal to
\beq
S_{2}^{Y}(\omega) = 1 + \eta_{1}\eta_{2}[S_{0}^{Y}(\omega)-1].
\eeq
It is impossible to measure this spectrum without disturbing
the feedback loop because all of the light in the $b_{2}$ beam must
be
incident upon the photodetector in order to measure $X_{2}$. However,
it is possible to measure the phase-quadrature of the out-of-loop beam
by homodyne detection. This was done in \cite{Tau95}, which
verified that this quadrature is also unaffected by the feedback, with
\beq
S_{3}^{Y}(\omega) = 1 + \eta_{1}\bar{\eta}_{2}[S_{0}^{Y}(\omega)-1].
\eeq

For simplicity, consider the case where the initial beam is coherent
with $S_{0}^{X}(\omega)=S_{0}^{Y}(\omega)=1$. Then
$S_{2}^{Y}(\omega)=1$ and
\beq \label{ltu}
S_{2}^{Y}(\omega)S_{2}^{X}(\omega) =
|1-{g} \tilde{h}(\omega) \exp(-i\omega T)|^{-2}.
\eeq
This can clearly be less than unity. This represents a violation of
the uncertainty relation (\ref{ur3}) which follows from the
commutation
relations (\ref{comr2}). In fact it is easy to show (as done first by
Shapiro
{\em et al.} \cite{Sha87}) from
the solution (\ref{X2fou}) that the commutation relations
(\ref{comr2}) are false for the field $b_{2}$ and must be replaced by
\beq
[\tilde{X}_{2}(\omega),\tilde{Y}_{2}(\omega')]
= \frac{4\pi i \delta (\omega + \omega') }
{1-{g} \tilde{h}(\omega) \exp(-i\omega T)}, \label{comr3}
\eeq
which explains how \erf{ltu} is possible.

At first sight, this apparent
violation of the canonical commutation relations
would seem to be a major problem of this theory. In fact, there are
no violations of the canonical commutation relations. As emphasized
in Sec.~\ref{sec:cq}, the canonical commutation relations
(\ref{341ccr2}) are
between fields at different points in space, {\em at the same time}.
It is only for free fields (traveling forward in space for an
indefinite time) that one can replace the space difference $z$ with a
time difference $t=z/c$. Field $b_{3}$ is such a free field, as it
can
be detected an arbitrarily large distance away from the apparatus.
Thus its quadratures at a particular point
do obey the time-difference commutation relations (\ref{comr1}), and
the corresponding Fourier domain relations (\ref{comr2}). But the
field $b_{2}$ cannot travel an indefinite distance before being
detected. The time from the second beam splitter to the detector
$\tau_{2}$ is a physical parameter in the feedback system.

For times shorter than the total feedback loop delay $T$ it can
be shown that
\beq
[X_{2}(t),Y_{2}(t')] = 2i\delta(t-t') \;\;\; {\rm for}\;\; |t-t'|<T.
\eeq
Now the field $b_{2}$ is only in existence for a time $\tau_{2}$
before it is detected. Because $\tau_{2} < T$, this means that the
time-difference commutation relations between different parts of field
$b_{2}$ are
actually preserved for any time such that those parts of the field
are in existence, traveling through space towards the detector. It is
only  at times greater than the
feedback loop delay time $T$ that non-standard commutation relations
hold. To summarize, the
commutation relations between any of the fields at different spatial
points always holds, but there is no reason to expect the time or
frequency commutation relations to hold for an in-loop field. Without
these relations, it is not clear how ``squeezing'' should be defined.
Indeed, it has been suggested \cite{Buc99} that ``squashed light''
would be a more appropriate term for in-loop ``squeezing'' because
the
uncertainty has actually been squashed, rather than squeezed out of
one
quadrature and into another.

\subsection{Semiclassical Theory} 
\label{Sec:sc}

A second theoretical reason against the use of the word squeezing to
describe the sub-shot-noise in-loop amplitude quadrature is that
(providing beam $b_{0}$ is not squeezed), the entire apparatus can be
described semiclassically. In a semiclassical description there is no
noise except classical noise in the field amplitudes, and shot noise
is a result of a quantum detector being driven by a classical beam of
light. That such a description exists
might seem surprising, given the importance of vacuum
fluctuations in the explanation of the spectra in Sec.~\ref{secvf}.
However, the semiclassical explanation, as explored in
\cite{Sha87,Tau95,KhoKil94}, is even simpler.

For the fields, let us use the same symbols as before, but with a cl
superscript to
remind us that these are classical variables rather than operators.
Then the only irreducible source of noise in the system is
the shot noise at the two detectors $k=2,3$. This arises 
from the assumption that a classical light field of frequency 
$\omega_{0}$ and time-flux of energy $W^{\rm cl}$ induces photo-electron 
emissions as a random process at rate $W^{\rm cl}/\hbar \omega_{0}$. 
Here $\hbar$ appears as a universal phenomenological constant relating
the output of photodetectors to the incoming light.
In addition to this irreducible noise, the light field itself may 
have (classical) noise which is represented by the amplitude fluctuation 
variable $X^{\rm cl}(t)$, which is scaled in the same way as the
quantum operator $X(t)$ has been.

In the linearized
approximation
we have been working in, the Poissonian shot-noise fluctuations can
be approximated by Gaussian fluctuations, giving
\beq
\delta I_{k}(t) = \sqrt{I_{k}}\,[X_{k}^{\rm cl}(t) + \xi_{k}(t)],
\eeq
where $I_{k}$ is the mean value of the photocurrent and $\xi_{k}(t)$
represents independent white-noise sources obeying
\beq
\ip{\xi_{k}(t')\xi_{j}(t)} = \delta_{jk}\delta(t-t').
\eeq
In our case we have $\sqrt{I_{2}} = \beta \sqrt{\eta_{1}\eta_{2}}$ and
$\sqrt{I_{3}}= \beta \sqrt{\eta_{1}\bar{\eta}_{2}}$

Using the above expression (\ref{etatb}) for $\delta \eta$, the 
fluctuation in the field
incident on detector 2 is
\bqa
X_{2}^{\rm cl}(t) &=& \sqrt{\eta_{1}\eta_{2}}\,X_{0}^{\rm
cl}(t-T_{2}) \nn \\
&&+\, {g}\int_{0}^{\infty}h(t') [X_{2}^{\rm cl}(t-T-t') +
\xi_{2}(t-T-t')]dt'.
\eqa
Assuming stable feedback (the conditions are the same as before) and
solving this in Fourier domain we get
\beq
\tilde{X}_{2}^{\rm cl}(\omega) = \frac{
\exp(-i\omega T_{2})\sqrt{\eta_{2}\eta_{1}}\,
\tilde{X}_{0}^{\rm cl}(\omega) +
{g} \tilde{h}(\omega) \exp(-i\omega T)\tilde{\xi}_{2}(\omega)
}{1-{g} \tilde{h}(\omega) \exp(-i\omega T)}.
\eeq
But the photocurrent noise itself is proportional to
\beq
\tilde{X}_{2}^{\rm cl}(\omega) + \tilde{\xi}_{2}(\omega)
 =
 \frac{\exp(-i\omega T_{2})\sqrt{\eta_{2}\eta_{1}}\,
 \tilde{X}_{0}^{\rm cl}(\omega)
 + \tilde{\xi}_{2}(\omega)}{1-{g} \tilde{h}(\omega) \exp(-i\omega T)}.
\eeq

From \erf{exptdet} we find the spectrum
\beq \label{S2Xa}
S_{2}^{X}(\omega) =
\frac{1+\eta_{2}\eta_{1}[S^{X}_{0}(\omega)-1]}{|1-{g}
\tilde{h}(\omega) \exp(-i\omega T)|^{2}},
\eeq
where I have defined
\beq
S_{0}^{X}(\omega) = 1 + \ip{X^{\rm cl}_{0}(0)\tilde{X}^{\rm
cl}_{0}(\omega)}
\eeq
as the spectrum which would be observed with no feedback and no beam
splitters. With this identification, the expression
\erf{S2Xa} is identical with \erf{S2X} derived using a quantized
field. 

In this formulation, the sub-shot noise of the in-loop current is no
surprise at all. It is simply a result of feeding back an amplified
negative version of the shot noise randomly produced at the detector
so that the current at a later time will be anticorrelated with
itself. The low noise is in the photocurrent only, not in the
light beams which here are all completely classical. This explains
why the out-of-loop detector does not produce a sub-shot noise signal.
The photocurrent noise at that detector is proportional to
\bqa
\xi_{3}(t) + X_{3}^{\rm cl}(t) &=&   \xi_{3}(t)  +
\sqrt{\eta_{1}\bar{\eta}_{2}}\,X_{0}^{\rm cl}(t-T_{2}) 
+ {g}\sqrt{\bar{\eta}_{2}/\eta_{2}} \nonumber \\
&&\times\,
\int_{0}^{\infty}h(t') [X_{2}^{\rm cl}(t-T-t') 
+ \xi_{2}(t-T-t')]dt'.
\eqa
Since $\xi_{3}$ is independent of everything else, it is noise which
cannot be reduced by the feedback. Only the classical noise
$X_{0}^{\rm cl}$ can be reduced, and that at the expense of
introducing the extra uncorrelated noise $\xi_{2}$. Calculating the
spectrum $S_{X}^{3}(\omega)$ again gives the same answer as the
quantum theory.

\subsection{QND Measurements of In-loop Beams}
\label{QNDinloop}

On the basis of the above semiclassical theory
it might be thought that all of the calculations of noise spectra
made in
this section relate only to the noise of photocurrents and say
nothing about the noise properties of the light beams themselves.
While this appears to be the case from the semiclassical theory, it
is not really true, as can be seen from considering quantum
non-demolition (QND) measurements \cite{WalMil94}. In this context,
a QND detector is one which can measure the intensity of light
without absorbing it. Such devices cannot be described by semiclassical
theory, which shows that this theory is not complete and hence cannot
be expected to provide the correct intuition about the state of the
light itself.

A specific model for a QND device will be considered in
Sec.~\ref{secQND}. Here we simply assume that a perfect QND device
can measure the amplitude quadrature $X$ of a continuum field without
disturbing it beyond the necessary back-action from the Heisenberg 
uncertainty principle. In other words, the
QND device should give a read-out at time $t$ which can be
represented by the operator $X(t)$. The correlations of this read-out
will thus reproduce the correlations of $X(t)$.
For a perfect QND measurement of $X_{2}$ and $X_{3}$, the spectrum
will reproduce those of the conventional (demolition) photodetectors
which measure these beams. This confirms that these detectors
(assumed perfect) are indeed recording the true quantum fluctuations
of the light impinging upon them.

What is more interesting is to consider a QND measurement on $X_{1}$.
That is because the set up in Fig.~\ref{wfig1} is equivalent (as 
mentioned above) to a set up without
the second beam splitter, but instead with an in-loop photodetector
with efficiency $\eta_{2}$.
In this version, the beams $b_{2}$ and
$b_{3}$ do not physically exist. Rather, $b_{1}$ is the in-loop beam
and $X_{2}$ is the operator for the noise in the photocurrent produced by the
detector. As shown above, this operator can have vanishing noise at
low frequencies for ${g} \to -\infty$. However, this is not
reflected in the noise in the in-loop beam, as recorded by our
hypothetical QND device. Following the methods of Sec.~\ref{ilols},
the spectrum of $X_{1}$ is
\beq
S_{1}^{X}(\omega) = \frac{1 + \eta_{1}[S_{0}(\omega)-1]
+ {g}^{2} |\tilde{h}(\omega)|^{2}\bar{\eta}_{2}/\eta_{2}}
{|1-{g} \tilde{h}(\omega) \exp(-i\omega T)|^{2}}.
\eeq
In the limit ${g} \to -\infty$, this becomes at low frequencies
\beq
S_{1}^{X}(0)  \to \frac{1 - {\eta}_{2}}{\eta_{2}},
\eeq
which is not zero for any detection efficiency $\eta_{2}$ finitely
less than one. Indeed, for $\eta_{2} < 0.5$ it is above shot-noise.

The reason that the in-loop amplitude quadrature spectrum is not
reduced to zero for large negative feedback is that the feedback loop
is feeding back noise $X_{\mu}(t)$ in the photocurrent fluctuation operator
$X_{2}(t)$ which is independent of the fluctuations in the amplitude
quadrature $X_{1}(t)$ of the in-loop light. The smaller $\eta_{2}$
is,
the larger the amount of extraneous noise in the photocurrent and the
larger the noise introduced into the in-loop light. In order to
minimize the low-frequency
noise in the in-loop light, there is an optimal feedback
gain. In the case $S_{0}(\omega)=1$ (a coherent input), this is given
by
\beq \label{goptfor1}
{g}_{\rm opt} = - \frac{\eta_{2}}{1-\eta_{2}},
\eeq
giving a minimum in-loop low-frequency noise spectrum
\beq \label{Sminfor1}
S_{1}^{X}(0)_{\rm min} = 1-\eta_{2}.
\eeq
The fact that the detection efficiency does matter in the attainable
in-loop squeezing shows that these are true quantum fluctuations.

\subsection{A Squeezed Input}
\label{secasi}

Although the feedback device discussed in this section cannot produce
a free squeezed beam, it is nevertheless useful for reducing
classical noise in the output beam $b_{3}$. It is easy to verify the
result of \cite{Tau95} that if one wishes to reduce classical
noise $S_{0}^{X}(\omega)-1$ at a particular frequency $\omega$, then
the optimum feedback is such that
\beq \label{optlam}
{g} \tilde{h}(\omega) \exp(-i\omega T) =
- \eta_{1}\eta_{2}[S_{0}^{X}(\omega)-1].
\eeq
This gives the lowest noise level in the amplitude of $b_{3}$ at that
frequency
\beq
S_{3}^{X}(\omega)_{\rm opt} = 1
+ \frac{\bar{\eta}_{2}\eta_{1}[S_{0}^{X}(\omega)-1]}{1 +
\eta_{2}\eta_{1}[S_{0}^{X}(\omega)-1]}.
\eeq

For large classical noise we have feedback proportional to
$S_{0}^{X}(\omega)$ and an optimal noise value of $1/\eta_{2}$, as
this
approaches the limit of \erf{limlam}. The interesting regime
\cite{Tau95} is the
opposite one, where $S_{0}^{X}(\omega) - 1$ is small, or even
negative.
This last case corresponds to to a squeezed input beam. Putting
squeezing through a beam splitter reduces the squeezing in both
output beams. In this case, with no feedback the residual squeezing
in beam
$b_{3}$ would be
\beq
S_{3}^{X}(\omega)_{\rm no} = 1 +
\bar{\eta}_{2}\eta_{1}[S_{0}^{X}(\omega)-1],
\eeq
which is closer to unity than  $S_{0}^{X}(\omega)$. The optimal
feedback (the purpose of which is to reduce noise) is, according to
\erf{optlam}, positive. That is to say, destabilizing feedback
actually puts back in beam $b_{3}$ some of the squeezing lost through 
the beam splitter. Since the
required round-loop gain (\ref{optlam}) is less than unity,
the feedback loop remains stable.

This result highlights the nonclassical nature of squeezed
fluctuations.
When an amplitude squeezed beam strikes a beam splitter, the
intensity
at one output port is anticorrelated with that at the other, hence
the need for positive feedback. Preliminary observations of this
effect were reported in~\cite{Tau96}.
Of course the feedback can never put more
squeezing into the beam than was present at the start. That is,
$S_{3}^{X}(\omega)_{\rm opt}$ always lies between $S_{0}^{X}(\omega)$
and $S_{3}^{X}(\omega)_{\rm no}$. However,
if we take the limit $\eta_{1} \to 1$ and $S_{0}^{X}(\omega) \to 0$
(perfect squeezing to begin with)
then all of this squeezing can be recovered, for any $\eta_{2}$. This
effect might even be of practical use for attenuating highly squeezed
sources, such as those produced by laser diodes, while retaining most
of the squeezing \cite{Tau96}.

\section{Quantum Langevin Equations}
\label{secqle}

In the preceding section, only continuum fields were considered, with
all optical and electro-optical devices being treated as classical
(i.e. deterministic) systems. Often one wishes to consider feedback
onto quantum
systems which may introduce extra noise terms into the equations. To
do this one needs a theory for describing the dynamics of localized
quantum systems interacting with continuum fields. In the optical
regime this theory was put on a rigorous foundation by Gardiner and
Collett \cite{GarCol85}. This theory is based on an electric-dipole
coupling, and involves key approximations which rely on a
rapid oscillation (at optical frequency $\omega_0$) of the system
dipole due to
the system Hamiltonian $H_0$. Let that dipole, in the interaction
picture of
$H_0$, be proportional to
\beq
c(t) \exp(-i\omega_0 t)+c\dg(t)\exp(i\omega_0t).
\eeq
Here $c$ and $c\dg$ are
dimensionless Heisenberg-picture system operators, normalized so that $c c\dg
\ket{0}=\ket{0}$,
where $\ket{0}$ is the ground state of $H_0$.

Let the system be located at $z=0$ and consider an incoming ($z<0$)
and outgoing ($z>0$) continuum field with operator $b(z,t)$ as
defined
previously. Then, under the rotating wave approximation \cite{Gar91},
the dipole coupling to the field at $z=0$ can be modeled by the
Hamiltonian
\beq V(t) = i\sqrt{\gamma}[c(t) b\dg(0,t) - b(0,t) c\dg(t)].
\label{coup} \eeq
Here $\hbar=1$
and $\gamma \ll \omega_{0}$ is the characteristic decay rate of the
system.
 For a general
interaction one would have to use both space and time arguments for
the continuum field. However, the coupling considered here is
strictly local.
Both before ($z<0$) and after ($z>0$) the interaction, the field
still freely
propagates. This means that, although it is necessary to use a
spatial as well
as a temporal argument, the spatial argument need only have two
values: {\em
before} and {\em after}. In order to conform with pre-established
usage
\cite{GarCol85}, these values will be called {\em in} and {\em out}.
The input field is defined (in the Heisenberg picture) as
\beq
b_{\rm in}(t) = b(0^-,t),
\eeq
and the output field as
\beq
b_{\rm out}(t) = b(0^+,t).
\eeq
Both of these fields obey the commutation relations
\beq \label{hcr}
[b(t),b\dg(t')]=\delta(t-t'),
\eeq
at least for time differences shorter than the delay time in any feedback
loop.

The interaction of the external field with the system at time $t$ is
described
by the Hamiltonian (\ref{coup}). It is necessary to be
careful in dealing with this because of the singular commutation
relations
(\ref{hcr}). A convenient way is to use the quantum \ito stochastic
calculus
\cite{GarCol85}. In the Heisenberg picture, the infinitesimal unitary
transformation pertaining to the interaction at time $t$ is
\beq \label{hint}
U(t,t+dt)=\exp\left[ \sqrt{\gamma}
c(t) dB_{\rm in}\dg(t) - \sqrt{\gamma}c(t)\dg dB_{\rm in}(t) 
-iH(t)dt\right],
\eeq
where $H$ is a system Hamiltonian representing any small perturbation 
on top of the free energy (such as classical driving), and
\beq
dB_{\rm in}(t) = b_{\rm in}(t)dt
\eeq
is the quantum analogue of the Wiener increment \cite{GarCol85}. Note
that the
field operator in the unitary transformation (\ref{hint}) is the bath
immediately before it interacts with the system, rather than the bath
at the
instant it is interacting with the system. This is the appropriate
operator if
one treats the stochastic increment $dB_{\rm in}(t)$ in the \ito
sense. This
means that \erf{hint} must be expanded to second order in the
increment. For an input in the vacuum state, the operator $dB_{\rm
in}$ has a vanishing first order moment, and a single nonvanishing
second-order
moment \cite{GarCol85}
\beq \label{sinsom}
dB_{\rm in}(t) dB_{\rm in}\dg(t) = dt.
\eeq
This could be thought of as vacuum noise.

Applying the unitary transformation (\ref{hint}) to an arbitrary
system operator
$s(t)$ yields
\bqa
s(t+dt) &=& U\dg(t,t+dt) s(t) U(t,t+dt) \nn \\
&=& s + \gamma \left( c\dg s c - \half c\dg c s - \half s c\dg c
\right)dt \nn \\
&& -\,[-i Hdt + \sqrt{\gamma}dB_{\rm in}\dg(t) c - \sqrt{\gamma}c\dg
dB_{\rm in}(t) , s],\label{QLE1}
\eqa
where all
system operators in \erf{QLE1}
have the time argument $t$. This is what I will refer to as a quantum
Langevin equation (QLE) for $s$. Because $dB_{\rm in}(t)$ is the bath
operator
before it interacts with the system, it is independent of the system
operator $s(t)$. Hence one can derive
\beq \label{meanval1}
\ip{\dot{s}} = \left\langle \left[\gamma\left( c\dg s c - \half c\dg c s -
\half s c\dg c \right) + i [H,s] \right]\right\rangle.
\eeq

From this it apparent that there is a \sch picture representation of
the system dynamics. In this picture
\beq
\ip{\dot{s}(t)} = {\rm Tr}[\dot{\rho}(t)s],
\eeq
where $\rho(t)$ is the state matrix for the system which evidently
obeys
\beq \label{me1}
\dot\rho = \gamma{\cal D}[c]\rho -i[H,\rho],
\eeq
where for arbitrary operators $A$ and $B$
\beq
{\cal D}[A]B = ABA\dg - \half A\dg A B - \half B A\dg A.
\eeq
An equation of the form (\ref{me1}) is known as a master equation. I
will discuss the master equation further in Sec.~\ref{sec:me}.

Although the noise terms in (\ref{QLE1}) do not contribute to
\erf{meanval1}, they  are necessary in order for \erf{QLE1}
to be a valid Heisenberg equation of motion. If
they are omitted then the commutation relations of the system will
not be preserved.
As well as giving the evolution of the system, (\ref{hint})
transforms the input field operator into the output field operator:
\beq
b_{\rm out}(t) = U\dg(t,t+dt) b_{\rm in}(t) U(t,t+dt) = b_{\rm in}(t)
+ \sqrt{\gamma}c(t).
\eeq
Assuming that $b_{\rm out}(t)$ has a large coherent amplitude then
one
can then derive an expression for its linearized intensity
fluctuations, which are proportional to
\beq
X_{\rm out}(t) = X_{\rm in}(t) + \sqrt{\gamma}\delta x(t),
\eeq
where $X_{\rm out}(t)$ is the fluctuation quadrature operator for the 
output field as usual, and where
$\delta x = x-\ip{x}_{\rm ss}$, where $x=c+c\dg$ is a system 
quadrature with mean steady-state value $\ip{x}_{\rm ss}$.

\section{Feedback based on Nonlinear Measurements}
\label{secfnm}

\subsection{QND Measurements}
\label{secQND}
Section~\ref{secils} showed that it was not possible to create
squeezed light in the conventional sense
using ordinary photodetection and linear feedback. Although the
quantum theory appeared to show that the light
which fell on the detector in the feedback loop was sub-shot noise,
this could not be extracted because it was demolished by the
detector. An obvious way around this would be to use a quantum
non-demolition quadrature (QND) detector.
One way to achieve this is for two fields of different frequency
to interact via a nonlinear refractive index. In order to obtain a
large nonlinearity, large intensities are required. It  is easiest to
build up large intensities by using a resonant cavity. To describe
this requires the theory of quantum Langevin equations just presented.

\begin{figure}
\includegraphics[width=1\textwidth]{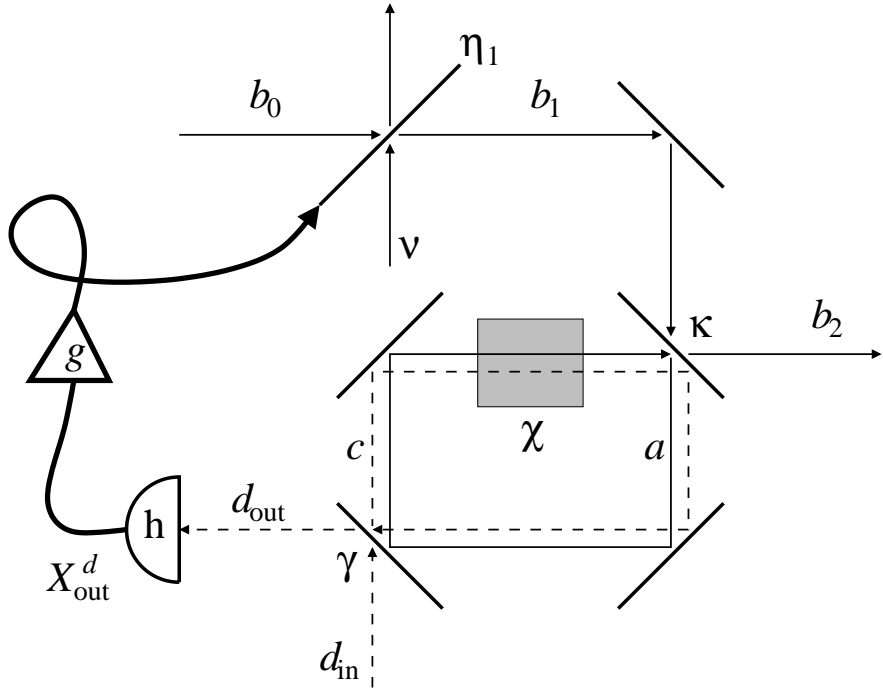}
	\caption{Diagram for a traveling-wave feedback experiment based on a
	QND measurement. Traveling
	fields are denoted $b$ and $d$. The first beam splitter
	transmittance $\eta_{1}$ is variable. A cavity (drawn as a ring cavity
	for convenience) supports two modes, $a$ (solid line) and $c$ (dashed
	line). The decay rates for these two modes are $\kappa$ and $\gamma$
	respectively. They are coupled by a nonlinear optical process
	indicated by the crystal labeled $\chi$.
	The perfect homodyne detection at the detector labeled h yields
	a photocurrent proportional to $X^{d}_{\rm out} = d_{\rm out}+d_{\rm
	out}\dg$ }
	\label{wfig2}
\end{figure}

Consider the apparatus shown in Fig.~\ref{wfig2}. The purpose of the
detector in the feedback loop
is to make a QND measurement of the quadrature $X^{b}_{\rm in}$ of
the field $b_{\rm in}$. This fields drives a cavity mode with decay
rate $\kappa$ described by annihilation operator $a$. This mode is
coupled to a second mode with annihilation operator $c$ and decay
rate $\gamma$.
Ignoring practicalities, I will take the coupling
between the two modes to be the ideal QND coupling, as considered
initially by Hillery and Scully \cite{HilScu82} and Yurke
\cite{Yur85}, namely
\beq \label{QNDHam}
H = \frac{\chi}{2}x^{a}y^{c},
\eeq
where
\beq
x^{a} = a+a\dg \;;\;\;\; y^{c} = -ic+ic\dg.
\eeq
As described in~\cite{AlsMilWal88}, this Hamiltonian could
in principle be realized by a crystal with a $\chi^{(2)}$
nonlinearity, combined with other processes.
This Hamiltonian commutes with the $x^{a}$ quadrature of mode $a$,
and causes this to drive the $x^{c}$ quadrature of mode $c$. Thus
measuring the $X^{d}_{\rm out}$
quadrature of the output field $d_{\rm out}$ from mode $c$ will give
a QND measurement of $a+a\dg$, which is approximately a QND
measurement of $X^{b}_{\rm in}$.

Following the analysis of~\cite{AlsMilWal88}, the quantum
Langevin equations for the quadrature operators are
\bqa
\dot{x}^{a} &=& -\frac{\kappa}{2}x^{a} - \sqrt\kappa\,X^{b}_{\rm in} ,\\
\dot{x}^{c} &=& -\frac{\gamma}{2}x^{c} - \sqrt\gamma\,X^{d}_{\rm in}
+ \chi x^{a}.
\eqa
These can be solved in the frequency domain as
\bqa
\tilde{x}^{a}(\omega) &=& -\frac{\sqrt\kappa\,\tilde{X}_{\rm
in}^{b}(\omega)}{\kappa/2 + i\omega} \label{xafou}\\
\tilde{x}^{c}(\omega) &=& -\frac{\sqrt\gamma\,\tilde{X}_{\rm
in}^{d}(\omega)+\sqrt{\kappa}\chi \tilde{X}_{\rm
in}^{b}(\omega)/(\kappa/2+i\omega)}{\gamma/2 + i\omega}.
\eqa
The quadrature of the output field $d_{\rm out} = d_{\rm in} +
\sqrt{\gamma}c$ is therefore
\beq
\tilde{X}^{d}_{\rm out}(\omega) = -\frac{\gamma\kappa Q
\,\tilde{X}_{\rm
in}^{b}(\omega)/(\kappa+2i\omega)+ (\gamma-2i\omega)\tilde{X}_{\rm
in}^{d}(\omega)}{\gamma  + 2i\omega},
\eeq
where I have defined a quality factor for the measurement
\beq
Q = 4\chi/\sqrt{\gamma\kappa}.
\eeq
In the limits $\omega \ll \kappa,\gamma$ and $Q \gg 1$
we have
\beq \label{gml}
\tilde X^{d}_{\rm out}(\omega) \simeq
-Q \tilde X_{\rm in}^{b}(\omega),
\eeq
which shows that a measurement (by homodyne detection) of the $X$ quadrature of
$d_{\rm out}$ does indeed
effect a measurement of the low frequency variation in
$X_{\rm in}^{b}$.

To see that this measurement is a QND measurement we have to calculate
the statistics of the output field from mode $a$, that is $b_{\rm
out}$. From the solution (\ref{xafou}) we find
\beq
\tilde{X}^{b}_{\rm out}(\omega) = -
\frac{\kappa/2-i\omega}{\kappa/2+i\omega}\tilde{X}^{b}_{\rm
in}(\omega).
\eeq
That is, apart from an irrelevant phase factor the output field is
identical to the input field, as required for a QND measurement. Of
course we cannot expect the other quadrature to remain unaffected,
because of the uncertainty principle. Indeed we find
\beq \label{Ybout}
\tilde{Y}^{b}_{\rm out}(\omega) =  -
\frac{(\kappa-2i\omega)}{\kappa  + 2i\omega}\tilde{Y}_{\rm
in}^{b}(\omega) +
\frac{Q \gamma\kappa \,\tilde{Y}_{\rm
in}^{d}(\omega)/(\gamma +2i\omega)}{\kappa  + 2i\omega} ,
\eeq
which shows that noise has been added to $Y_{\rm in}$. Indeed, in the
good measurement limit which gave the result (\ref{gml}) we find the
phase quadrature output to be dominated by noise:
\beq
\tilde{Y}^{b}_{\rm out}(\omega) \simeq
Q \tilde{Y}_{\rm in}^{d}(\omega).
\eeq

\subsection{QND-based Feedback}

We know wish to show how a QND measurement, such as that just
considered, can be used to produce squeezing via feedback. The
physical details of how the feedback can be achieved are as outlined
in Sec.~\ref{secdd} In particular, in the limit that the transmittance
$\eta_{1}$ of the modulated beam splitter goes to unity, the effect
of the modulation is simply to add an arbitrary signal to the amplitude
quadrature of the controlled beam. That is, the modulated beam can be
taken to be
\beq \label{2s0}
b_{1}(t) = b_{0}(t-\tau_{1}) + \beta \delta\eta_{1}(t-\tau_{1}),
\eeq
where $b_{0}$ is the beam incident on the modulated beam splitter, as
in Sec.~\ref{secdd}. In the present case, $b_{1}(t)$ is then fed into the QND
device, as shown in Fig.~\ref{wfig2}, so
\beq \label{2s1}
b_{\rm in}(t) = b_{1}(t),
\eeq
and the modulation is controlled by the photocurrent from a (assumed
perfect) homodyne
measurement of $X^{d}_{\rm out}$:
\beq \label{2s2}
\delta\eta_{1}(t) = \frac{{g} }{-Q\beta}
\int_{0}^{\infty} h(s) X^{d}_{\rm out}(t-\tau_{0} - s) ds.
\eeq
Here $\tau_{0}$ is the delay time in the feedback loop, including the
time of flight from the cavity for mode $c$ to the homodyne detector,
and $h(s)$ is as before.

Substituting~(\ref{2s0})-(\ref{2s2}) into the results of the
preceding subsection we find
\beq
\tilde{X}_{\rm out}^{b}(\omega) = -
\frac{\kappa -2i\omega}{\kappa +2i\omega}
\frac{\tilde{X}_{\rm in}^{b}(\omega)+
\tilde{X}_{\rm in}^{d}(\omega){g}\tilde{h}(\omega)
Q^{-1}
\frac{\gamma -2i\omega}{\gamma +2i\omega}}
{1-{g} \tilde{p}(\omega)\tilde{h}(\omega)\exp(-i\omega T)},
\eeq
where $T=\tau_{1}+\tau_{0}$ is the total round-trip delay time, and
\beq
\tilde{p}(\omega) = \frac{\gamma\kappa}
{(\kappa+2i\omega)(\gamma+2i\omega)}
\eeq
represents the frequency response of the two cavities. If we assume
that the field $d_{\rm in}$ is in the vacuum state then we can
evaluate the spectrum of amplitude fluctuations in $X_{\rm out}^{b}$
to be
\beq
S_{\rm out}^{X}(\omega) =
\frac{S_{\rm in}^{X}(\omega)+{g}^{2}Q^{-2}}{
|1-{g} \tilde{p}(\omega)\tilde{h}(\omega)\exp(-i\omega T)|^{2}}.
\eeq

Clearly in the limit $Q \to \infty$,
the added noise term in the amplitude spectrum
can be ignored. Then for sufficiently large negative ${g}$ the
feedback will
produce a sub-shot noise spectrum. Note the difference between this
case and that of Sec.~\ref{secils}. Here the squeezed light is not part of the
feedback loop; it is a free beam.  The ultimate limit to how squeezed
the beam can be is caused by the noise in the measurement. In the
limit ${g} \to -\infty$ we find
\beq
S_{\rm out}^{X}(\omega)_{\rm min} =
|Q\tilde{p}(\omega)\tilde{h}(\omega)|^{-2}.
\eeq

Since $b_{\rm out}$ is a free field, not part of any feedback loop,
it should obey the standard commutation relations. This is the case, as
can be verified from the expression (\ref{Ybout})
for $Y_{\rm out}^{b}$ (which is
unaffected by the feedback). Consequently, the spectrum for the
phase quadrature
\beq
S_{\rm out}^{Y}(\omega) =
S_{\rm in}^{Y}(\omega) +
|Q\tilde{p}(\omega)|^{2}
\eeq
shows the expected increase in noise.  It is not difficult to show
that the uncertainty product $S_{\rm out}^{Y}(\omega)S_{\rm out}^{X}(\omega)$
is greater than or equal to unity, as
required.
\label{secqbf}

\subsection{Parametric Down Conversion}

The preceding section showed that feedback based on a perfect QND
measurement can produce squeezing. This has never been done
experimentally because of the difficulty of building a perfect QND
measurement
apparatus. However, it turns out that QND measurements are not the
only way to produce squeezing via feedback. Any mechanism which
produces correlations between the beam of interest and another beam
which
are more ``quantum'' than the correlations between the two outputs of
a linear beam splitter can be the basis for producing squeezing via
feedback. Such a mechanism must involve some sort of optical
nonlinearity, which is why this section V is entitled  ``Feedback
based on Nonlinear Measurements''.

The production of amplitude squeezing  by feeding back a
quantum - correlated signal was predicted \cite{JakWal85} and observed
\cite{WalJak85b} by Walker and Jakeman in 1985, using the process of
parametric down-conversion. An improved feedback scheme for the same
system was
later used by Tapster, Rarity and Satchell \cite{TapRarSat88}
to obtain an inferred amplitude spectrum $S_{1}^{X}(\omega)_{\rm
min} = 0.72$ over a limited frequency range.

The essential idea is as follows. A non-degenerate optical parametric
oscillator
(NDOPO) can be realized by pumping a crystal with a $\chi^{(2)}$
nonlinearity by a laser of frequency $\omega_{1}$ and momentum ${\bf
k}_{1}$. For particular choices of $\omega_{1}$ the crystal can be
aligned so that a pump photon can be transformed into
 a pair of photons each with frequencies $\omega_{2},\omega_{3}$ and
 momenta ${\bf k}_{2},{\bf k}_{3}$ with
 $\omega_{1}=\omega_{2}+\omega_{3}$ and ${\bf k}_{1} = {\bf
 k}_{2}+{\bf k}_{3}$.
A pair of down-converted photons will thus be correlated in time
because they are
produced from a single pump photon. On a more macroscopic level, this
means that the amplitude quadrature
fluctuations in the two down-converted beams are
positively correlated, and the coefficient of correlation can in
principle be very close to unity. In this ideal limit,
measuring the intensity of beam 2
should give a readout identical to that obtained from measuring beam
3.In effect, the measurement of beam 2 is like a QND measurement of
beam 3. Thus, feeding back the photocurrent from beam 2 with an
negative gain should
be able to reduce the noise in beam 3 below the shot-noise limit. In
the limit of perfect detection, the optimum gain becomes arbitrarily
large.

In ~\cite{JakWal85,TapRarSat88} the negative feedback was
effected by controlling the power of the pump laser (which controls
the rate at which photon pairs are produced). This maintains the
symmetry of the experiment so that the fed-back photocurrent from
beam $2$ has the same statistics as the photocurrent from the free
beam $3$. In other words, they will both be below the shot-noise,
whereas without feedback they are both at best shot-noise limited.
However, it is not necessary to preserve the symmetry in this way.
The measured photocurrent from beam $2$ can be fed {\em forward} to
control the amplitude fluctuations in beam $3$ (for example by using
an electro-optic modulator as described in Sec.~\ref{secdd}). This
feedforward was realized experimentally
by Mertz {\em et al.} in 1990 \cite{Mer90} achieving
similar results to that obtained by feedback \cite{MerHeiFab91}. The
lesson is that unless one is concerned with light inside a feedback
loop, there is no essential difference between feedback and
feedforward.
Indeed, the squeezing produced by QND-based feedback discussed in
Sec.~\ref{secqbf} could equally well have
been produced by QND-based feedforward.

\subsection{Second Harmonic Generation}

A third example of using quantum correlations between two beams to
good effect in a feedback (or feedforward) loop is that of
second-harmonic generation. Like parametric down conversion, this
involves a $\chi^{(2)}$ crystal, but here the two low frequency modes
are assumed to be degenerate, so that $\omega_{2}=\omega_{3}$ and
${\bf k}_{2}={\bf k}_{3}$, and the polarizations are the same also.
For simplicity, I will call the low frequency mode the red mode and
the high frequency mode with frequency $\omega_{1} = 2\omega_{2}$ the
green mode. For second harmonic generation, it is the red mode which
is pumped (that is, the reverse of parametric down-conversion). In
this case, the pumped mode (red) is also treated as a quantum system
so that the two beams are the red and green beams. The system is
easily treated using quantum Langevin equations \cite{ColLev91}.

The proposal made in~\cite{WisTauBac95} is as follows.
The first harmonic (red) resides in a good cavity and
is driven by a laser at the end with very high reflectivity.
The second harmonic (green), which is generated in the red cavity, is
reflected at one end only (giving two passes)
so that it forms a single output beam. By itself
(that is, in the absence of feedback) this system produces amplitude
squeezing in both the first and second harmonic \cite{Pas94}.
The amplitude squeezing
in the red can be understood to be due to two-photon
absorption, which is the effect of the adiabatically eliminated green
mode. Such nonlinear absorption preferentially damps large
fluctuations and so reduces the variance.
The antibunching in the green mode can be attributed to the fact
that the creation of a green photon requires the loss of two red
photons, which reduces the chances of this event re-occurring.

Without feedback, the optimum low frequency squeezing in the red mode
output is
\beq
S^{X}_{r}(0) = \frac{2}{3} \;\;\; {\rm for} \;\; \chi = \frac{1}{3},
\label{nofbr}
\eeq
and that in the green mode is
\beq
S_{g}^{X}(0) = \frac{1}{9} \;\;\; {\rm for} \;\; \chi \to \infty.
\label{nofbg}
\eeq
Here $\chi$ is the ratio of the nonlinear intensity loss rate to the
linear
intensity loss rate for the red mode.
Now there are two possible ways to feed back onto the driving of
the system. The first is to use the green photocurrent to control the
amplitude of the red driving, in order to
try to reduce the noise in the red output light. A linearized
treatment
gives a new minimum of
\beq
S^{X}_{r}(0)  = \frac{1}{2} \;\;\; {\rm for} \;\; \chi = 1, \;\;
g = \frac{1}{2},
\eeq
which is significantly less than the no-feedback value (\ref{nofbr}).
Here,
$g$ is the low frequency loop transfer function,
equal to the round-loop gain
of the feedback \cite{Ste90}. Note that it is positive, corresponding
to destabilizing feedback. This is different from the cases of QND-based
feedback and down-conversion-based feedback.

The other sort of feedback is to control the driving of the red mode
using the detected red photocurrent, looking to enhance the squeezing
in
the green output. It turns out that the minimum $S_{g}^{X}(0)$
of~(\ref{nofbg}) as
$\chi \to \infty$ cannot be lowered by feedback. However, for finite
values of $\chi$, the feedback can give a definite improvement. For
example, with no feedback
\beq
S_{g}^{X}(0) = \frac{1}{2} \;\;\; {\rm for} \;\; \chi = 1,
\eeq
whereas the optimal feedback gives
\beq
S_{g}^{X}(0)  = \frac{1}{3} \;\;\; {\rm for} \;\; \chi = 1, \;\;
g = \frac{1}{4}.
\eeq

All of these results are calculated for the case of
unit efficiency photodetectors. The effect of non-unit efficiency is to
reduce the effectiveness of the feedback, and to alter the conditions
of optimality. The general solution, including the full
spectrum with an arbitrary transfer function $g \tilde{h}(\omega)$,
is contained in~\cite{WisTauBac95}.

\section{Quantum Trajectories}
\label{secqt}

\subsection{The Master Equation}\label{sec:me}

The Quantum Langevin Equation (QLE) considered in Sec.~\ref{secqle}
are Heisenberg picture equations which detail the effect of the bath
on the system. They also, through the input-output relations, specify
the effect of the system on the bath, and hence the relation between
the system evolution and the measured photocurrents. For systems
which have linear QLEs, this is the easiest method to analyze their
evolution. However, even some simple systems (such as a two-level
atom) do not have linear QLEs. Also, sometimes a better intuition
about feedback can be gained by working in the \sch picture rather
than the \hei picture.

If one is interested only in the evolution of the system then there
is a simple \sch picture equivalent to the QLE. This is the quantum
master
equation, an example of which was already derived as \erf{me1} from
the corresponding QLE. Unlike a QLE, a master equation is always
a linear equation. It is possible to derive the master equation
directly from the system-bath coupling without deriving the QLE,
simply by tracing over the bath. I present this derivation below,
because it shows clearly that this step (ignoring the bath) is not an
essential one. If instead of ignoring the bath, one measures it, then
one obtains a quantum trajectory equation. This is a different sort
of \sch picture equivalent to the QLE which is more general than the
master equation, as it can also relate the system
evolution to the measured photocurrents. This is necessary if one is
to consider feeding back the photocurrent to alter the dynamics of the
system.

For convenience, I will work in the interaction
picture, rather than the \sch picture, so that the free
evolution causing the oscillation of the dipole can be ignored.
 The interaction between the system and the bath is now given by
\begin{equation}
	V(t) = {i}\hbar \sqrt{\gamma}\,[ b\dg(t) c  - c\dg b(t)],
\end{equation}
where $c$ is a slowly-varying interaction picture operator, and
$b(t)$
is also a slowly-varying interaction picture bath operator. The
time-dependence is maintained for the bath operator, because the free
Hamiltonian of the bath causes propagation at the speed of light, so
a new part of the bath interacts with the system at each new point in
time.
These parts are labeled by the time of interaction $t$. Thus, $b(t)$
is an operator in the Hilbert space for a particular part of the
bath. Each part has its own state matrix $\mu(t)$. For the incoming
 field to be a bath requires that its total state matrix be the
 direct product of the state matrices of the parts
\cite{GarParZol92}.
 That is to say, the temporally separate parts of the bath must be
 unentangled.

Let the system at time $t$ be known to be $\rho(t)$. Thus the initial
state of the system and (relevant part of the) bath at time $t$ is
\begin{equation}
	R(t)=\mu(t)\otimes\rho(t).
\end{equation}
The infinitesimally evolved state is
\begin{equation}
	R(t+dt)=U_{}(t,t+dt)[\mu(t)\otimes\rho(t)]U_{}\dg(t,t+dt),
\end{equation}
where
\begin{equation}
U_{}(t,t+dt) = \exp \left[ \sqrt{\gamma}\,dB_{}^\dagger (t)
c -  c\dg \sqrt{\gamma}\,dB_{} (t)  -iHdt \right] , \label{351U1}
\end{equation}
where $H$ is the system Hamiltonian in the interaction picture.
If the input bath is in the vacuum state then all of the first and
second order moments of $dB(t)$ vanish except, as in \erf{sinsom},
\beq
{\rm Tr}[dB_{}(t)dB\dg_{}(t) \mu(t)] = dt.
\eeq
Thus, it is necessary to
expand some of the terms in $U(t,t+dt)$ to second order.
The result for $R(t+dt)$ is
\begin{eqnarray}
     &  & \mu(t)\otimes\rho(t) + \sqrt{\gamma}\,\left[ dB_{}^\dagger
(t) c - c\dg dB_{} (t) , \mu(t)\otimes\rho(t) \right] \label{351tot}
\\
	 &&  +\gamma \left\{dB_{}\dg(t) \mu(t) dB_{}(t) \otimes c\rho(t) c\dg
- \smallfrac{1}{2} dB_{}(t)dB_{}\dg(t) \mu(t) \otimes c\dg c
\rho\right. \nn \\
&&  - \left. \smallfrac{1}{2} \mu(t) dB_{}(t)dB_{}\dg(t) \otimes \rho c\dg
	 c\right\} -i[Hdt,\mu(t)\otimes\rho(t)].\nn
\end{eqnarray}
The infinitesimally evolved reduced state matrix for the system is
given by
\begin{equation}
	\rho(t+dt)={\rm Tr}_\mu [ R(t+dt)].
\end{equation}
Taking the trace over the bath state in~(\ref{351tot}) yields the
master equation (\ref{me1}):
\beq \label{411av1}
\dot\rho = \gamma{\cal D}[c]\rho -i[H,\rho].
\eeq
The first term here represents irreversible evolution (damping in 
this case), and is of the 
unique form for such evolution as proved by Lindblad \cite{Lin76}. 
The second term represents unitary evolution which is of course 
reversible.

\subsection{Photon Counting}

To consider measuring the output bath it is useful to explicitly
write down its state. Before interacting with the system,
the input bath state is $\mu(t) = \ket{0}\bra{0}$ for all time $t$,
where
$\ket{0}$ is the lowest eigenstate for $a\dg a$. Here, $a =
\sqrt{dt}\,b_0(t) = dB_0(t)/\sqrt{dt}$,
so that $a\dg a$ has integer
eigenvalues representing the number of photons arriving in the
interval
of time $[t,t+dt)$.  Let the state of the system at time $t$ be
$\rho(t)$,
independent of $\mu(t)$. The entangled state after the interaction of
duration $dt$ is, from~(\ref{351tot})
\begin{eqnarray}
	R(t+dt)&=&\ket{0}\bra{0}\otimes\rho(t) + \sqrt{\gamma dt} \left[
	\ket{1}\bra{0}\otimes c\rho(t) + \ket{0}\bra{1}\otimes\rho(t)c\dg
	\right] \nn \\
	&& - {i} dt \ket{0}\bra{0}\otimes[H,\rho(t)]
	+ \;\gamma dt \left\{ \ket{1}\bra{1}\otimes c\rho(t)c\dg\right.
	\nn \\
	&& - \left. \half \ket{0}\bra{0}\otimes \left[c\dg c \rho(t) +
	\rho(t) c\dg c \right] \right\}. \label{411ent}
\end{eqnarray}
The free dynamics of the electromagnetic field will now remove the
bath state from the system, so that the entanglement in \erf{411ent} 
will be maintained. If
the outgoing bath is ignored, then the system obeys the master equation
(\ref{411av1}).

 To obtain information about the system, the outgoing field
must be measured. The obvious measurement to consider is photon
counting.
This I will model by projecting the field states onto eigenstates of
$a\dg a$. It is possible to consider specific models for photon
detectors, usually based on atomic systems \cite{Gar91}. However,
these
simply remove the measurement step, where possibilities become
actualities, one step further along the von Neumann chain
\cite{Von32}.
There is little which results from such models which cannot be
achieved by
including losses and convolving the classical photocurrent with an
empirically derived detector response function. Therefore, I will
simply
use projection measurement operators
\begin{equation}
	P_0 = \ket{0}\bra{0} \; ; \;\; P_1 = \ket{1}\bra{1}.
	\label{411proj}
\end{equation}

If there is a null count then
the unnormalized state matrix of the system and bath
 (whose norm gives the probability of this outcome)
 is given by
 \beq
 \tilde{R}_{0}(t+dt) = P_{0}R(t+dt)P_{0} = P_{0}\otimes
\tilde{\rho}_{0}(t+dt),
\eeq
where
\begin{equation}
	\tilde{\rho}_0(t+dt) = \rho(t) -  \gamma \smallfrac{1}{2}\{c\dg c,
\rho(t)\}
       dt - {i}[H,\rho(t)]dt.
\end{equation}
This has a norm only infinitesimally different from one, so
for almost all time intervals
no photons are detected in the output.

If a photon is detected, then $\tilde{R}_{1}(t+dt) = P_{1} \otimes
\tilde{\rho}_{1}(t+dt)$. That is, the system jumps into the
unnormalized
conditioned state
\begin{equation}
	\tilde{\rho}_1(t+dt) = \gamma c\rho(t) c\dg dt.
\end{equation}
The norm of this state matrix is equal to the probability of a
detection occurring in the interval $[t,t+dt)$,
and is equal to $\gamma \langle c\dg c\rangle dt$.
Clearly, the unconditioned master equation evolution (\ref{411av1})
is
retrieved by averaging over the two possible results
\begin{equation}
	\rho(t+dt)=\tilde{\rho}_0(t+dt)+\tilde{\rho}_1(t+dt).
\end{equation}

It is useful and elegant
to reformulate this evolution in the form of an explicitly stochastic
evolution equation. This equation of motion specifies the quantum
trajectory of the system. Since the measurement result is a point
process, it can be represented by a random variable $dN_{\rm c}(t)$
representing the increment (either zero or one)
in the photon count in the interval $[t,t+dt)$. It is formally
defined by
\begin{eqnarray}
	{\rm E}[dN_{\rm c}(t)] & = & {\rm Tr}[c\dg c \rho_{\rm c}(t)]\gamma
dt,
	\label{412dN1} \\
	dN_{\rm c}(t)^2 & = & dN_{\rm c}(t).
	\label{412dN2}
\end{eqnarray}
Here the subscript ${\rm c}$ indicates that the quantity to which it
is attached
is conditioned on previous measurement results, arbitrarily far back
in time.
The conditioned state matrix obeys the
stochastic master equation (SME)
\begin{equation}
\label{412me3}
d\rho_{\rm c}(t) = \left\{ dN_{\rm c}(t) {\cal G}[c] - dt {\cal
H}\left[{i}H +
\smallfrac{1}{2}\gamma c^\dagger c\right] \right\} \rho_{\rm c}(t).
\end{equation}
Here, the nonlinear (in $\rho$) superoperators ${\cal G}$ and ${\cal
H}$ are
defined  by
\begin{eqnarray}
{\cal G}[r]\rho &=& \frac{r \rho r^\dagger }{{\rm Tr}[r \rho
r^\dagger ]} - \rho , \label{defcalG} \\
{\cal H}[r]\rho &=& r\rho + \rho r^\dagger - {\rm Tr}[r\rho+\rho
r^\dagger]\rho . \label{defcalH}
\end{eqnarray}
The nonlinearity of the SME (\ref{412me3}) is indicative of the
fundamental nonlinearity of quantum measurements. The original master
equation for $\rho(t)={\rm E}[\rho_{\rm c}(t)]$
can be restored simply by replacing $dN_{\rm c}(t)$ in~(\ref{412me3})
by its ensemble average value (\ref{412dN1}).

Because of the assumed perfect detection, the stochastic equation
for the state matrix is equivalent to a stochastic equation
for the state vector. The unraveling of the master equation into a
stochastic \sch equation
representing photon counting is
the most commonly used quantum trajectory for numerical simulations
\cite{DalCasMol92,MolCasDal93,DumZolRit92,DumParZolGar92,TiaCar92,%
MarDumTaiZol93}. From the point of view of
measurement theory, the stochastic master equation is of more use
than the stochastic \sch equation because
it is more transparently related to the unconditioned master equation
and because it can be generalized to cope with inefficient detectors.
This will be dealt with explicitly for the case of homodyne detection.

\subsection{Homodyne Detection Theory}

The unraveling of the master equation (\ref{411av1}) as a quantum
trajectory is not unique. Different detection schemes will result in
different quantum trajectory equations. For squeezing, the most
useful detection technique is homodyne detection.
In the simplest configuration, the output field of the cavity,
$b_{\rm out}=\nu + \sqrt{\gamma}c$, is sent through a beam
splitter of transmittance $\eta$ very close to one. Into the other input
port of the
beam splitter is injected a very strong coherent field. This has the same
frequency as the system dipole, and is known as the local
oscillator. The transmitted field is then represented by the operator
\begin{equation}
	b_1 = \nu + \sqrt{\gamma}c + \beta,
	\label{421pluslo}
\end{equation}
where $\beta$ is a complex number representing a coherent amplitude,
such that $|\beta|^2/(1-\eta)$ is equal to the input photon flux of
the
local oscillator. The photodetection operators are then applied as
above, with the annihilation operator defined as $a =
b_{1}\sqrt{dt}$.

Let the coherent field $\beta$ be real, so that the homodyne
detection
leads to a measurement of the $x$ quadrature of the system dipole.
Also, let us measure time in units of $\gamma^{-1}$ so that this
parameter disappears from our equations. Then
the rate of photodetections at the (perfect) detector
\begin{equation}
	{\rm E}[dN_{\rm c}(t)]={\rm Tr}[(\beta^2 + \beta x + c\dg
c)\rho_{\rm c}(t)]dt.
\end{equation}
where $x=c+c\dg$ as previously.
In the limit that $\beta$ is much larger than $c$, this rate consists
of
a large constant term plus a term proportional to $x$, plus a small
term.
It is not difficult to show that the SME
for the conditioned state matrix is altered from \erf{412me3} to
\begin{equation}
	d\rho_{\rm c}(t) = \left\{ dN_{\rm c}(t){\cal G}[c+\beta] + dt{\cal
H}
	\left[-{i}H -\beta c - \smallfrac{1}{2} c\dg c \right]\right\}
	 \rho_{\rm c}(t).
	\label{421sme1}
\end{equation}

The ideal limit of homodyne detection is when the local oscillator
amplitude goes to infinity. In this limit, the rate of
photodetections
goes to infinity, but the effect of each on the system goes to zero,
because the field being detected is almost entirely due to the local
oscillator. Thus, it should be possible to approximate the
photocurrent
by a continuous function of time, and also to derive a smooth
evolution
equation for the system. This was done first by Carmichael
\cite{Car93b}.
A  more rigorous working of the derivation is found in~\cite{WisMil93a}.
The result is that over a time much longer
than $\beta^{-3/2}$, but much smaller than unity, the
system evolution can be approximated by the SME
\begin{equation}
	d\rho_{\rm c}(t)=-{i}[H,\rho_{\rm c}(t)]dt+{\cal D}[c]\rho_{\rm
c}(t)dt
	 + dW(t) {\cal H}[c]\rho_{\rm c}(t).
	\label{422sme3}
\end{equation}
Here $dW(t)$ is an infinitesimal Wiener increment \cite{Gar85}
satisfying
\bqa
{\rm E}[dW(t)]&=& 0,\\
dW(t)^{2} &=& dt.
\eqa
Thus, the jump evolution of~(\ref{421sme1}) has been
replaced by diffusive evolution. Equation~(\ref{422sme3}) is, by its
derivation,  an It\^o stochastic master equation, where the
equal-time stochastic increment $dW(t)$ is independent of the state
of the
system $\rho_{\rm c}(t)$. It is trivial to see that the ensemble
average evolution reproduces the nonselective master equation
(\ref{411av1}) by eliminating the noise term.

Just as the $\beta \to \infty$ leads to continuous evolution for the
state, it also changes the point process photocount
into a continuous photocurrent with white noise. Removing the
constant local-oscillator contribution and scaling appropriately gives
\begin{equation}
	I_{\rm c}^{\rm hom}(t) \equiv \lim_{\delta t \to
0}\lim_{\beta\to\infty}
	\frac{\delta N_{\rm c}(t) - \beta^2\delta t}{\beta \delta t}  =
	\ip{x}_{\rm c}(t) + 	\xi(t),
	\label{422hompc}
\end{equation}
where $\xi(t)=dW(t)/dt$.
It is not difficult to see that if the detector efficiency is $\eta$,
the homodyne photocurrent becomes
\begin{equation}
	I_{\rm c}^{\rm hom}(t) \equiv \lim_{\delta t \to
0}\lim_{\beta\to\infty}
	\frac{\delta N_{\rm c}(t) - \eta\beta^2\delta t}{\sqrt{\eta}\,\beta
\delta t}
	= \sqrt{\eta}\, \ip{x}_{\rm c}(t) + \xi(t),
	\label{422hompc2}
\end{equation}
and the SME (\ref{422sme3}) is modified to
\begin{equation}
	d{\rho}=-{i}[H,\rho_{\rm c}(t)]dt+{\cal D}[c]\rho_{\rm c}(t)dt +
	\sqrt{\eta} dW(t)
	{\cal H}[c]\rho_{\rm c}(t).
	\label{63sme1}
\end{equation}

\subsection{Homodyne-mediated Feedback}

In the following section I will consider the use of feedback to
produce or enhance squeezing in intracavity fields.
Since squeezing is the reduction in fluctuations of one quadrature of
a field, the obvious sort of feedback to consider is one using the
homodyne photocurrent obtained from measuring the output field. Here
I will
develop the theory for describing this sort of feedback. As well as
being more relevant for our purposes than feedback using the direct
detection photocurrent \cite{Wis94a}, it is also somewhat easier to
treat theoretically, which is why it was derived first
\cite{WisMil93b,WisMil94a}.

In principle, the homodyne photocurrent could be subject to any sort
of filtering prior to being fed back, including nonlinear filtering.
However it turns out that for the applications we wish to consider,
only linear filtering is desired. Also, rather than using a response
function $h(t)$ I will simply take the feedback to be delayed by a
time $T$. That means that the evolution due to the feedback can
simply
be written as
\beq \label{63fb1}
[\dot{\rho}_{\rm c}(t)]_{\rm fb} = I_{\rm c}^{\rm hom}(t-T)
{\cal K}\rho_{\rm c}(t)/\sqrt{\eta},
\eeq
where ${\cal K}$ is a superoperator. Since $I_{\rm c}^{\rm hom}(t)$
may be negative, ${\cal K}$ must be such as to give valid
evolution
irrespective of the sign of time. That is to say, it must give
reversible
evolution with
\beq \label{defF}
{\cal K}\rho \equiv -{i}[F,\rho]
\eeq
for some Hermitian operator $F$. In other words, we can represent the 
effect of the feedback by the Hamiltonian
\beq \label{Hfb}
H_{\rm fb} = F\left[\ip{c+c\dg}_{\rm c}(t-T) +\xi(t-T)/\sqrt{\eta}\right].
\eeq

Because the stochasticity in the measurement (\ref{63sme1}) and the
feedback (\ref{63fb1}) is Gaussian white noise, it is relatively
simple to determine the effect of the feedback.
Bearing in mind that the feedback must act after the
measurement, and that~(\ref{63fb1}) must be interpreted as 
 a Stratonovich equation 
\cite{WisMil94a},
the result for the total conditioned evolution of the system is
\begin{eqnarray}
\rho_{\rm c}(t+dt) &=& \left\{1+ {\cal K}
[\langle c+c^\dagger\rangle_{\rm c}(t-T)dt
+ dW(t-T)/\sqrt{\eta}\,] + \frac{1}{2\eta} {\cal K}^2dt \right\}
\nonumber \\
&& \times\bigl\{ 1 + {\cal H}[-{i}H]dt + {\cal D}[c]dt +
\sqrt{\eta}dW(t)
{\cal H}[c] \bigr\}\rho_{\rm c}(t). \label{63sme2}
\end{eqnarray}
For $T$ finite, this becomes
\begin{eqnarray}
	d\rho_{\rm c}(t) & = & dt\left\{ {\cal H}[-{i}H] + {\cal D}[c] +
	\ip{c+c\dg}_{\rm c}(t-T) {\cal K} + \frac{1}{2\eta}{\cal K}^2
	\right\}\rho_{\rm c}(t)
	\nonumber \\
	 &  & + \, dW(t-T){\cal K}\rho_{\rm c}(t)/\sqrt{\eta} +
	 \sqrt{\eta}dW(t){\cal H}[c]
	 \rho_{\rm c}(t).
\end{eqnarray}
On the other hand, putting $T=0$ in \erf{63sme2} gives
\begin{eqnarray}
	d\rho_{\rm c}(t) & = & dt\left\{ -{i}[H,\rho_{\rm c}(t)] +
	 {\cal D}[c]\rho_{\rm c}(t)  - 	{i}[F,c\rho_{\rm c}(t) +
	 \rho_{\rm c}(t) c\dg ]\right. \nn \\
	 &  & +\, \left. {\cal D}[F]\rho_{\rm c}(t)/\eta \right\}
	+ dW(t){\cal H}[\sqrt{\eta}c-{i}F/\sqrt{\eta}\,]\rho_{\rm c}(t).
	\label{63sme3}
\end{eqnarray}

\label{sec:hfb}

\section{Intracavity Squeezing}
\label{secics}

\subsection{The Linear System}

In order to understand the effect of quantum limited feedback on
intracavity squeezing, it is
useful to consider an exactly solvable system. In this
section, I will mainly be following~\cite{WisMil94a} in
considering the case of a linear optical system, with linear
feedback based on homodyne (or QND) detection.
By a linear system, I mean that the
equation of motion for the two quadrature operators are linear. This
is approximately the case for many quantum optical systems, in the limit
of large photon numbers. For specificity, I will chose a system which is
exactly linear. If, as in the remainder of this chapter, one is
interested in the behaviour of one quadrature only (here the $x$ quadrature),
then all linear dynamics can be composed of damping, driving, and parametric
driving. Damping will be assumed to be always present (as necessary to do
feedback or obtain an output from the cavity) and will have rate $1$.
Constant linear driving simply shifts the origin away from $x=0$, and will
be ignored. Stochastic linear driving in the white noise approximation
causes diffusion in the $x$ quadrature, at a rate $l$. Finally, if the
strength of the parametric driving ($H \sim xy$) is $\theta$ (where $\theta=1$
would represent a degenerate parametric oscillator at threshold), then the
master equation for the system is
\beq \label{64me1}
\dot{\rho} = {\cal D}[a]\rho + \smallfrac{1}{4}l{\cal D}[a\dg-a]\rho
+
\smallfrac{1}{4}\theta[a^2 - a\dg{}^2,\rho] \equiv {\cal L}_0\rho,
\eeq
where $a$ is the annihilation operator for the cavity mode.

An alternative definition for the linearity of the $x$ quadrature
dynamics is that the marginal distribution of the Wigner function for
$x$
(which is the true probability distribution for $x$) obeys an
Ornstein-Uhlenbeck equation. That is to say,
\beq \label{OUW1}
\dot{P}(x) = \left( \partial_x k x + \smallfrac{1}{2} D \partial_x^2
\right) P(x),
\eeq
where $k$ and $D$ are constants.
The solution of this equation is a Gaussian with variance
\beq
V = \frac{D}{2k}.
\eeq
For the particular master equation above (the properties of which
will be
denoted by the subscript $0$), the drift and diffusion constants are
\begin{eqnarray}
	k_0 & = & \smallfrac{1}{2} (1+\theta),\label{64k1}
	\\
	D_0 & = & 1+l.
\end{eqnarray}
In this case, $V_0=(1+l)/(1+\theta)$.
If this is less than unity, the system exhibits squeezing of $x$. It
is
more useful to work with the normally ordered variance, which becomes
negative if the $x$ quadrature is squeezed. Here, I will denote it
\beq
U = V - 1,
\eeq
which for this system takes the value
\beq \label{ng4}
U_0 = \frac{l-\theta}{1+\theta}.
\eeq
If the system is to stay below threshold (so that the $y$ quadrature
does
not become unbounded), then the maximum value for $\theta$ is one. At
this
value, $U_0=-1/2$ when the $x$ diffusion rate $l=0$. Therefore
 the minimum value of squeezing which this linear system can
attain as a stationary value is half of the theoretical minimum of
$U=-1$.

\subsection{Homodyne-Mediated Feedback}

We now wish to consider the effect of homodyne-mediated
feedback on the intracavity
light. This is most easily understood using the quantum trajectory
picture in the Markovian limit. Thus we want the stochastic master
equation for the conditioned
state matrix $\rho_{\rm c}(t)$ \erf{63sme3}
\bqa \label{smeg}
d{\rho}_{\rm c}(t) &=&  dt\left( {\cal L}_0\rho_{\rm c}(t) +
{\cal K} [a\rho_{\rm c}(t) +\rho_{\rm c}(t) a\dg ]
+ \frac{1}{2\eta} {\cal K}^2 \rho_{\rm c}(t) \right) \nn \\
&&+\, dW(t)\left( \sqrt{\eta}{\cal H}[a] + {\cal K}/\sqrt{\eta}\right)
\rho_{\rm c}(t).
\eqa
Here, ${\cal L}_0$ is as defined in~(\ref{64me1}).

The question now arises as to what to choose for the ${\cal K}$.
Seeking to reduce the fluctuations in $x$ suggests the feedback
operator, which is related to ${\cal K}$ by \erf{defF}, should be
\beq \label{64F}
F = -\lambda y / 2.
\eeq
As a separate Hamiltonian, this translates a state in the negative
$x$ direction for $\lambda$ positive. By controlling this Hamiltonian by
the homodyne photocurrent
one thus has the ability to change the statistics for $x$ and
perhaps achieve better squeezing. This Hamiltonian can be effected by
driving the cavity (at a second mirror which can be assumed to have a
negligible loss rate compared to the first mirror).
Using this choice and changing \erf{smeg}
into a stochastic Liouville equation for the
conditioned Wigner function gives
\begin{eqnarray}
d{P}_{\rm c}(x) &=&  dt \left[ \partial_x (k_0+\lambda) x +
\frac{1}{2} \partial_x^2
\left( D_0 + 2\lambda + \lambda^2/\eta \right) \right] P_{\rm c}(x)
\nonumber \\
&& + \;  dW(t)\left[  \sqrt{\eta} \left( x-
\bar{x}_{\rm c}(t)  +  \partial_x \right) +
(\lambda/\sqrt{\eta}) \partial_x  \right] P_{\rm c}(x),
\end{eqnarray}
where $\bar{x}_{\rm c}(t)$ is the  mean of the distribution
$P_{\rm c}(x)$ and $dW(t)$ is as usual.

This equation is obviously no longer a simple Ornstein-%
Uhlenbeck equation. Nevertheless, it still has a Gaussian as
an exact solution, as can be shown by direct substitution.
The mean
$\bar{x}_{\rm c}$  and variance $V_{\rm c}$  of the
conditioned Gaussian distribution are found to obey
\begin{eqnarray}
\dot{\bar{x}}_{\rm c} &=& -(k_0+\lambda)\bar{x}_{\rm c} + \xi(t)
\left[
\sqrt{\eta} \left( V_{\rm c} - 1 \right) -
(\lambda/\sqrt{\eta}) \right], \label{dbx} \\
\dot{V}_{\rm c} &=& -2k_0 V_{\rm c} + D_0 - \eta \left(V_{\rm c} - 1
\right)^2. \label{dV}
\end{eqnarray}
Two points about the evolution equation for $V_{\rm c}$ are worth
noting. It is completely deterministic (no noise terms), and
it is not influenced by the presence of feedback.
Furthermore, for this linear system, it is
independent of $\bar{x}_{\rm c}$. Thus, the  stochasticity and
feedback terms in the equation for the mean do not even
enter that for the variance indirectly.

The equation for the conditioned variance is more simply
written in terms of the conditioned normally ordered variance
 $U_{\rm c} = V_{\rm c} - 1$
\begin{equation} \label{condU}
\dot{U_{\rm c}} = -2k_0 U_{\rm c} - 2k_0+D_0 - \eta U_{\rm c}^2.
\end{equation}
On a time scale as short as a cavity lifetime, $U_{\rm c}$ will
approach its stable steady-state value of
\begin{equation}  \label{Qcss}
U_{\rm c} = \eta^{-1} \left( -k_0 + \sqrt{k_0^2 + \eta (-2k_0+D_0)}.
\right)
\end{equation}
Substituting the steady-state conditioned variance into~(\ref{dbx}) gives
\begin{equation}
\dot{\bar{x}}_{\rm c} = -(k_0+\lambda)\bar{x}_{\rm c} +
\xi(t)\frac{1}{\sqrt{\eta}} \left[ -k_0
+ \sqrt{k_0^2 + \eta (-2k_0+D_0)} - \lambda \right]. \label{dbx2}
\end{equation}
If one were to choose
\beq \label{optlam3}
\lambda = -k_0 + \sqrt{k_0^2 + \eta (-2k_0+D_0)}
= -k_0 + \sqrt{k_0^2 + 2\eta k_0 U_0}
\eeq
then there would be no noise at all in the conditioned mean and so
one could set $\bar{x}_{\rm c} = 0$. In other words, this value of
$\lambda$ is precisely the value required to
minimize the unconditioned variance under feedback. When all
fluctuations in the mean are suppressed, the
unconditioned variance is equal to
the conditioned variance.

In general, the unconditioned variance will consist of two
terms, the conditioned quantum variance in $x$ plus the
classical (ensemble) average variance in the conditioned
mean of $x$:
\begin{equation}
U_\lambda = U_{\rm c} + {\rm E} [\bar{x}_{\rm c}^2 ].
\end{equation}
The latter term is found from~(\ref{dbx2}) to be
\begin{equation}
{\rm E} [\bar{x}_{\rm c}^2 ] = \eta^{-1} \frac{1}{2(k_0+\lambda)}
\left[ - (k_0+\lambda)
+ \sqrt{k_0^2 + \eta(-2k_0+D_0)} \right]^2.
\end{equation}
Adding (\ref{Qcss}) gives
\begin{equation}
U_\lambda = \eta^{-1} \frac{\lambda^2 +
\eta(-2k_0+D_0)}{2(k_0+\lambda)}= (k_0+\lambda)^{-1}\left( k_0 U_0 +
	\frac{\lambda^2}{2\eta}\right).
\end{equation}

An immediate consequence of this expression is that $U_\lambda$ can
only
be negative if $U_0$ is. That is to say, classical
feedback based on homodyne detection cannot {\em produce}
intracavity squeezing.
However, this does not mean
that the feedback cannot {\em enhance} squeezing.
Obviously, the best intracavity squeezing will be when
$\eta=1$, in which case the intracavity squeezing can be simply
expressed as
\begin{equation} \label{ng11}
U_{\rm min} = k_0 \left( -1 + \sqrt{1+R_0} \right),
\end{equation}
where $R_{0} = (-2k_0+D_0)/k_{0}^{2} \geq -1$. It can be proven that
that $U_{\rm min} \le U_0$, with equality
only if $\eta = 0$ or $U_0 = 0$.  This result implies
that the intracavity variance in $x$ can always be
reduced by classical homodyne-mediated feedback, unless it
is at the classical minimum. In particular, intracavity
squeezing can always be enhanced. For the parametric oscillator
defined
originally in~(\ref{64me1}), with $l=0$, $U_{\rm min} =
-\theta/\eta$. For $\eta = 1$, the (symmetrically ordered) $x$ variance
is
$V_{\rm min}=1-\theta$. The $y$ variance, which is unaffected by
feedback, is seen from~(\ref{64me1}) to be $(1-\theta)^{-1}$.
Thus, with perfect detection, it is possible to
produce a minimum uncertainty squeezed state with
arbitrarily high squeezing as $\theta \rightarrow 1$. This
is not unexpected as a parametric amplifier (in an undamped
cavity) also produces minimum uncertainty squeezed states.
The feedback removes the noise which was added
by the damping which is necessary to do the measurement used
in the feedback.

The reason that this feedback cannot produce squeezing is that the
conditioning of the variance according to \erf{condU}
cannot change the sign of the normally-ordered variance $U$.
 The homodyne
measurement does reduces the conditioned variance,
except when it is equal to the classical minimum of 1.
The more efficient the measurement, the greater the
reduction. Ordinarily, this reduced variance is not
evident because the measurement gives a random shift to the
conditional mean of $x$, with the
 randomness arising from the shot noise
of the photocurrent. By appropriately feeding
back this photocurrent, it is possible to precisely
counteract this shift and thus observe the conditioned variance.

If the time delay $T$ in the feedback loop is not
negligible then the counteraction will be less than perfect. It is
possible to calculate this effect exactly for an arbitrary linear
feedback response using the quantum trajectory theory \cite{WisMil94a}.
However, it would
generally be easier to return to the approach based on
quantum Langevin equations \cite{GioTomVit99}. For short delay $T \ll 1$,
there is a
simple expression for the modified normally ordered variance:
\beq
U_{\lambda;T} = U_{\lambda}(1+\lambda T).
\eeq
For squeezed systems, with $U_0 < 0$, the optimum value of
$U_\lambda$
occurs for $\lambda$ negative, as shown above. Thus, the time
delay reduces the total squeezing by the factor
$(1+\lambda T)$. On the other hand, classical noise is reduced to
$U_\lambda > 0$ with $\lambda$ positive, so that the total noise is
increased by the factor $(1+\lambda T)$. Overall, the time delay
degrades the effectiveness of the feedback, as expected.

Note that the optimal $\lambda$ of \erf{optlam3}
has the same sign as $U_0$. That
is to say, if the system produces squeezed light, then the
best way to enhance the squeezing is to add a force which
displaces the state in the direction of the difference between
 the measured photocurrent and the desired mean photocurrent.
This is the opposite of what would be expected classically,
and can be attributed to the effect of homodyne measurement on
squeezed
states.  For
classical statistics ($U \ge 0$) , a higher than
average photocurrent reading [$\xi(t) > 0$] leads to the
conditioned mean $\bar{x}_{\rm c}$ increasing (except if $U=0$ in
which
case the measurement has no effect). However, for nonclassical
states with $U < 0$, the classical intuition fails as a
positive photocurrent fluctuation causes $\bar{x}_{\rm c}$ to
decrease.
This explains the counterintuitive negative value of $\lambda$
required in squeezed systems, which naively would be
thought to destabilize the system and increase fluctuations. 
The value of the positive feedback required (\ref{optlam3}) is such that
the overall restoring force $k_0+\lambda$ is still positive.

Succinctly, one can state that conditioning can be made
practical  by feedback. The intracavity noise reduction produced by
classical
feedback can be precisely as good as that produced by
conditioning. This reinforces the simple explanation as to why
homodyne-mediated classical feedback cannot produce
nonclassical states: because homodyne detection cannot.
Nonclassical feedback
(such as using the photocurrent to influence nonlinear
intracavity elements) may produce nonclassical states, but
such elements can produce nonclassical states without
feedback, so this is hardly surprising. In order to produce
nonclassical
states by classical feedback, it would be necessary to have a
nonclassical measurement scheme. That is to say, one which does not
rely on measurement of the extracavity light to procure information
about
the intracavity state. Intracavity measurements (in particular,
quantum non-demolition measurements) are not limited by the random
process of damping to the external continuum. The extra
term which the measurement introduces into the nonselective
master equation will not produce nonclassical states, but
may allow the measurement to produce nonclassical {\em
conditioned} states. One would thus expect that intracavity QND
measurements would enable feedback to overcomes the
classical limit, and I will now show that this is indeed the case.
\label{sechmf}

\subsection{QND-Mediated Feedback}

The natural choice of quantum non-demolition variable is the
quadrature to be squeezed, say $x$ as before.
I use the same model for a QND measurement as in
Sec.~\ref{secQND}. Mode $a$ is coupled to mode $c$ by the Hamiltonian
in \erf{QNDHam}. The other
dynamics of mode $a$ are defined as before by
its Liouville superoperator ${\cal L}_0$. The density operator for
both modes thus obeys the following master equation:
\begin{equation}\dot{R} = {\cal L}_0 R - \frac{\chi}{2}[x(c-c\dg),R]
+ \gamma{\cal D}[c]R. \label{66me2}
\end{equation}

In order to treat mode $c$ as part of the apparatus rather than part
of the system, it is necessary to eliminate its dynamics. This can be
done by assuming that it is heavily damped, with $\gamma$ much
larger than all other rates.
Then, apart from initial transients, it will have few photons
and will be slaved to mode $a$. Following standard techniques for
adiabatic elimination \cite{WisMil93a}
gives the master equation for $\rho$, the
density operator for mode $a$ alone, as
\begin{equation} \label{qndme6}
\dot{\rho} = {\cal L}_0\rho + \Gamma {\cal D}[x/2] \rho,
\end{equation}
where the measurement strength parameter is $\Gamma = 4\chi^2 /
\gamma.$

Now add homodyne measurement of the $b$ mode with
efficiency $\eta$. Starting from the conditioned state matrix $R_{\rm
c}$ before the
adiabatic elimination and following it through gives the
conditioning master equation for $\rho_{\rm c}$ \cite{WisMil94a}
\begin{equation}
d{\rho}_{\rm c} = dt{\cal L}_0\rho_{\rm c} + dt
\Gamma {\cal D}[x/2]\rho_{\rm c} +
\sqrt{\rm \eta\Gamma} dW(t) {\cal H}[x/2]\rho_{\rm c}.
\end{equation}
Normalizing the homodyne photocurrent so that the noise is
the same as in preceding sections gives
\beq \label{66pc}
I_{\rm c}(t) = \sqrt{\rm H}\ip{x}_{\rm c}(t) + \xi(t).
\eeq
Here I am using H (a capital $\eta$) for $\eta\Gamma$ as the
effective
efficiency of the measurement. This is related to the parameter $Q$
in Sec.~\ref{secQND} by ${\rm H}=\eta Q^{2}/4$.
Note that this is not bounded above by
unity, since it is possible for $\chi^2/\gamma$ to be much greater than
one even with $\chi$ much less than $\gamma$. Recall that all rates are
measured in units of the $a$ mode linewidth (which was $\kappa$ in
Sec.~\ref{secQND}).

The photocurrent (\ref{66pc}) can be used in feedback onto the
$a$ mode just as in preceding sections. A feedback term of the form
\begin{equation}
[\dot{\rho}_{\rm c}]_{\rm fb} =  I_{\rm c}(t-T) {\cal K}
\rho_{\rm c}/\sqrt{\rm H}
\end{equation}
gives, in the limit $T \rightarrow 0$, the conditioned evolution
\begin{eqnarray}
d{\rho}_{\rm c} &=& dt\left({\cal L}_0\rho_{\rm c}+ \Gamma
{\cal D}[x/2]\rho_{\rm c}+ {\cal K}\smallfrac{1}{2}[ x\rho_{\rm c}+
\rho_{\rm c} x ] + \frac{1}{2{\rm H}} {\cal K}^2
\rho_{\rm c} \right)
\nonumber \\
&& + \; dW(t) \left( \sqrt{\rm H} {\cal H}[x/2] + {\cal
K}/\sqrt{\rm H} \right) \rho_{\rm c}.
\end{eqnarray}

Using the same expressions as in Sec.~\ref{sechmf} implies that
the probability distribution for the $x$ quadrature obeys
\begin{eqnarray}
d{P}_{\rm c}(x) &=&  dt\left[ \partial_x (k_0+\lambda) x +
\frac{1}{2} \partial_x^2 \left( D_0  +
\lambda^2/{\rm H} \right) \right] P_{\rm c}(x) \nn \\
&& + \; dW(t) \left[ \sqrt{\rm H}
[x-\bar{x}_{\rm c}(t) ]  + (\lambda/\sqrt{H}) \partial_x \right]
P_{\rm c}(x). \label{qndwsel}
\end{eqnarray}
 The mean and variance of this conditioned distribution obey
\begin{eqnarray}
\dot{\bar{x}}_{\rm c} &=& -(k_0+\lambda)\bar{x}_{\rm c} + \xi(t)
\left(
\sqrt{\rm H} V_{\rm c}  - \lambda/\sqrt{\rm H} \right)  ,
\label{dbxqnd}\\
\dot{V}_{\rm c} &=& -2k_0V_{\rm c} + D_0 - {\rm H} V_{\rm c}^2 .
\label{dVqnd}
\end{eqnarray}
These equations are identical to the corresponding equations for
homodyne mediated feedback (\ref{dbx}) and (\ref{dV}) apart from the
replacement of $(V_{\rm c} - 1)$ by $V_{\rm c}$ and $\eta$ by ${\rm
H}$ in the measurement terms.
In the limit ${\rm H} \rightarrow \infty$,~(\ref{dVqnd}) predicts
an arbitrarily small steady-state conditioned variance. This is
characteristic of a good QND measurement. Of course, the quantum
noise has not been eliminated but rather redistributed. For ${\rm H}$
to be large requires $\Gamma$ to be large also,
so that the variance in the unsqueezed quadrature is greatly
increased by the measurement term in~(\ref{qndme6}). This
ensures that Heisenberg's uncertainty principle is not violated.

For this QND measurement the stationary value for $V$ from
\erf{dVqnd} is
\beq \label{scVqnd}
V_{\rm c} = {\rm H}^{-1}\left(-k_{0}+\sqrt{k_{0}^{2}+{\rm H}
D_{0}}\right).
\eeq
Thus choosing the feedback strength to be
\begin{equation}
\lambda = -k_0 + \sqrt{k_0^2 + {\rm H}D_0}. \label{lminqnd}
\end{equation}
eliminates the stochastic element in~(\ref{dbxqnd}). In this case,
the stationary conditioned
variance \erf{scVqnd} is the minimum achievable variance.
In the limit ${\rm H} \rightarrow \infty$,
it is easy to see that $V_{\rm min}=V_{\rm c}$ approaches the
theoretical minimum value of $0$. That is, perfect squeezing can
be produced inside the cavity by QND mediated feedback. In this
limit, one requires the feedback to be
very strong, with $\lambda \simeq \sqrt{{\rm H}D_0}$. Unlike the
homodyne mediated feedback case, $\lambda$ should always be
positive, as in accord with classical intuition. Indeed, all of the
features of QND mediated feedback conform to a classical theory
of feedback with measurements of finite accuracy (related to ${\rm
H}$). The quantum nature of the feedback is manifest only in the
increased fluctuations in $y$ due to the measurement back-action
not present classically.

\subsection{Mimicking a Squeezed Bath}\label{TomVit}

The application of feedback based on a QND homodyne
measurement to a cavity state  has also been considered by Tombesi and Vitali
\cite{TomVit94}. However, rather than directly trying to minimize the
$x$ quadrature variance of the field mode, as I have discussed above,
their goal was to mimic the dynamics produced by shining a broad-band squeezed
vacuum onto the cavity mirror. Being broad-band compared to the cavity
mode, a squeezed vacuum input is parametrized by two numbers, $N,M$
which change the single nonzero second-order moments of \erf{sinsom} to
\bqa
dB_{\rm in}\dg dB_{\rm in} &=& Ndt\;;\;\; dB_{\rm in} dB_{\rm in}\dg =
(N+1)dt\;;\;\; \nonumber \\
dB_{\rm in}dB_{\rm in} &=& (dB_{\rm in}\dg dB_{\rm in}\dg)^{*} = Mdt.
\eqa
Positivity of the bath state matrix requires $|M|^{2} \leq N(N+1)$
\cite{Gar91}. The
master equation resulting from such bath correlations is \cite{Gar91}
\bqa
\dot{\rho} &=& -i[H,\rho] + (N+1){\cal D}[a]+N{\cal D}[a\dg]\nonumber
\\
& &- M\half[a\dg,[a\dg,\rho]] - M^{*}\half[a,[a,\rho]].
\eqa
Tombesi and Vitali show that an equation of this form can be produced
using feedback based on a QND homodyne measurement and that, not
surprisingly, it can can produce intracavity squeezing.

\subsection{The Micromaser}

A final example of feedback onto an intracavity state which can give
nonclassical noise reduction
is that of the micromaser \cite{FilJavMey86}.
This consists of a small microwave cavity
through which a monochromatic beam of resonant two-level atoms
is passed. The atomic state upon exit can be measured and the result
used
in feedback. The case of modifying the cavity quality factor
has been considered by Liebman and Milburn \cite{LieMil95}.
Because of the nature of the
Jaynes-Cummings Hamiltonian, the micromaser dynamics are complicated
without feedback, and even more complicated with. However, one result
is easy to explain. In the limit of short transit time the atoms
(assumed all to enter in the upper state) act simply as a linear
amplifier of the
cavity mode. In the absence of feedback the stationary state is
thermal, with a photon number variance much greater than the mean.
With weak feedback, increasing the cavity damping rate whenever an
outgoing atom is
detected in the lower state, the photon distribution can be made
sub-Poissonian, with a variance equal to half the mean. For longer
transit times, the no-feedback dynamics show the effects of trapping
states (where the atom undergoes an integer number of Rabi cycles in
transit \cite{FilJavMey86}),
and the minimum stationary variance is typically as low as one
quarter of
the mean. Feedback can produce an arbitrary small minimum variance
near a trapping state. However, this result is very sensitive to the
transit
time and so may be washed out by a realistic atomic velocity profile.

\section{Feedback Master Equation}

\label{secfme}
In \erf{63sme3} we have the stochastic master equation for a system
undergoing homodyne measurement, with instantaneous linear feedback
of the homodyne photocurrent. This is a Markovian \ito stochastic
equation,  so that is possible to take the ensemble
average simply by removing the stochastic term. This removes all
nonlinear terms from the stochastic master equation, and gives the
deterministic master equation
\beq \label{63me2}
\dot{\rho}=-{i}[H,\rho] + {\cal D}[c]\rho
-{i}[F,c\rho + \rho c\dg] + \frac{1}{\eta}{\cal D}[F]\rho.
\eeq
I will call this the homodyne feedback master equation. The
first feedback term, linear in $F$, is the desired effect of the
feedback
which would dominate in the classical regime. The second feedback
term
causes diffusion in the variable conjugate to $F$. It can be
attributed
to the inevitable introduction of noise by the measurement step in
the
quantum-limited feedback loop. The lower the efficiency, the more
noise
introduced.

The homodyne feedback master equation can be rewritten in the
Lindblad
form \cite{Lin76} as
\begin{equation}
\dot{\rho} = -{i}\left[H + \smallfrac{1}{2}(c^\dagger F + F c) ,\rho
\right]+
{\cal D}[c -{i}F] \rho +
\frac{1-\eta}{\eta}{\cal D}[F]\rho \equiv {\cal L}\rho.
\label{homofbme}
\end{equation}
In this arrangement, the effect of the feedback
is seen to replace $c$ by $c-{i}F$, and to add an extra term to the
Hamiltonian, plus an extra diffusion term which vanishes for perfect
detection. It is possible to derive an analogous feedback master
equation
for direct detection, but this is not needed for squeezing.
In the limit where intensity fluctuations can be linearized as
amplitude quadrature fluctuations, direct detection is essentially
equivalent to homodyne detection of the amplitude quadrature.

It is very important to note that although \erf{homofbme} has the
appearance of a normal master equation, one cannot simply use it in
the customary way for calculating the spectrum of the photocurrent
used in the feedback loop. For example the two-time correlation
function of the in-loop photocurrent is not given by the standard
expression:
\beq
{\rm E}[I_{\rm c}^{\rm hom}(t+\tau)I_{\rm c}^{\rm hom}(t) ]
\neq \eta \ip{:x(t+\tau)x(t):} + \delta(\tau) ,
\eeq
where the normally
ordered two-time correlation function for $x$ is defined as
\beq
\ip{:x(t+\tau)x(t):} = {\rm Tr} \left\{ (c+c^\dagger)
e^{{\cal L}\tau} [c\rho(t) + \rho(t) c\dg ] \right\}.
\eeq
That is because the photocurrent at time $t$ changes the system
through the feedback as well as through the conditioning. Taking this
into account, it is not difficult to show \cite{Wis94a}
that the correct expression is
\begin{eqnarray} \label{ttcf}
{\rm E}[I_{\rm c}^{\rm hom}(t+\tau)I_{\rm c}^{\rm hom}(t) ] &=&
\eta {\rm Tr} \left\{ (c+c^\dagger)
e^{{\cal L}\tau} [(c-{i}F/\eta)\rho(t)\right. \nonumber \\
&+& \left. \rho(t) (c\dg+{i}F/\eta) ] \right\} + \delta(\tau).
\end{eqnarray}
Note that the feedback is present in both the term in square brackets
and the evolution by ${\cal L}$ for the time $\tau$. The former
presence means that
the in-loop photocurrent may have a sub-shot-noise spectrum, even if
the system dynamics are classical. That is to say, for an optical system 
it may be possible to use an a semiclassical analysis 
(with classical light fields and detector shot noise)
which nevertheless correctly predicts a sub-shot noise photocurrent 
spectrum. This is not surprising given the analogous result for free 
squeezing in Sec.~\ref{Sec:sc}.

\section{In-loop Squeezing Revisited}

\subsection{In-Loop Squeezing}

In Sec.~\ref{secils} it was shown that a sub-shot-noise in-loop
photocurrent is not evidence for squeezing of the in-loop light in
the usual sense, not least because the usual two-time
commutation relations do not apply to an in-loop field.
Also, a linear optical element (a beam splitter) fails to extract
any squeezing from the loop, and in fact produces an above-shot-noise
output. Nevertheless it was also shown that an in-loop nonlinear
optical element (a QND intensity meter) agrees with the intensity
statistics seen by the in-loop detector. This means that it is an
open question as to whether the effect of the in-loop ``squashed''
\cite{Buc99} light on other
nonlinear optical elements is more akin to that of squeezed light or
light with classical fluctuations.

The simplest nonlinear optical element is a two-level atom.
Shortly after the first observation of squeezing, Gardiner
\cite{Gar86} made a seminal prediction regarding
its effects on such an atom \cite{FicDru97}, namely that
 immersing an atom in broad-band squeezed light would break the
equality between the transverse decay rates for the two quadratures
of the atomic dipole.
In particular, one decay rate could be made arbitrarily small,
producing an arbitrarily narrow line in the power spectrum of the
atom's fluorescence.  This was seen, as the title of~\cite{Gar86}
proclaims, as a ``direct effect of squeezing''.

In this section I present work published recently \cite{Wis98b},
which considers the question of whether this atomic
line-narrowing is characteristic only of squeezing in the
conventional sense (`real squeezing'), or whether it can be produced
by light which
gives rise to a below shot-noise photocurrent by virtue of being
part of a feedback loop (`in-loop squeezing').  The answer is that
in-loop squeezing  {\em can} do the job. In fact,
the inhibition of the decay of one quadrature depends on the amount
of squeezing and the quality of mode-matching to the atom in
exactly the same way for in-loop squeezing  as for free-field
squeezing. Thus in-loop squeezing appears
likely to be an important tool for future experimental investigation
of the effect of low-noise light on atoms, as it is usually
easier to generate than free squeezing.

\begin{figure}
\includegraphics[width=1\textwidth]{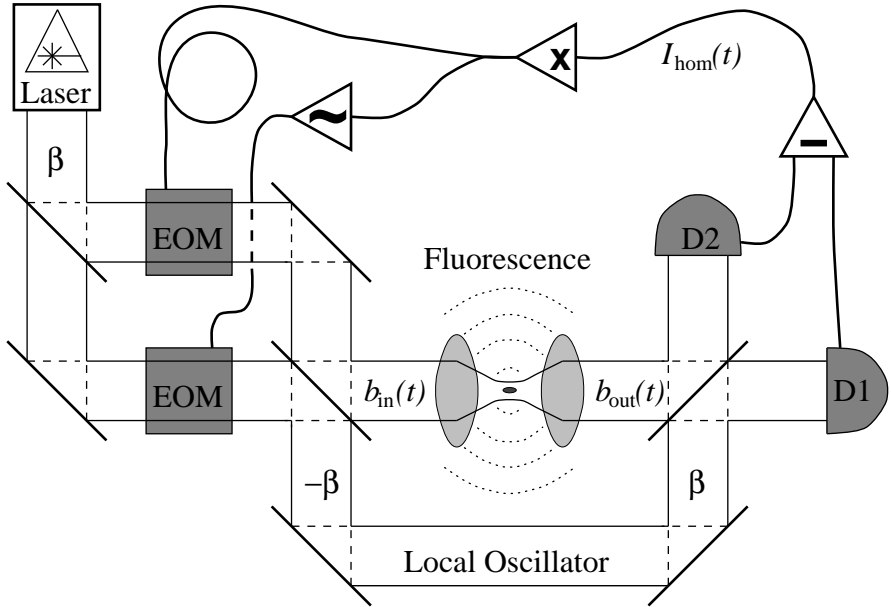}
	\caption{Diagram of the experimental configuration
	discussed in the text. All beam splitters are 50:50.
	The atom is represented by the small ellipse at
	the focus of $b_{\rm in}(t)$. The difference $I_{\rm hom}(t)$
	between the photocurrents at detectors D1 and D2 is  amplified and
	split. The two signals (with opposite sign) are
	fed back to the two electro-optic  	modulators (EOM)}
	\protect\label{wfig3}
\end{figure}

Consider the apparatus  shown in Fig.~\ref{wfig3}, but for the moment
without the fluorescent atom.
The Mach-Zehnder interferometer on the left
hand side has two functions. First it produces a weak beam
($b_{\rm in}$) which is given by
\beq
 b_{\rm in}(t) = \nu(t) -(i/2)(\beta e^{i\phi} - \beta e^{-i\phi})
\approx \nu(t) + \beta \phi.
\eeq
Here $\beta$, assumed real, is the coherent amplitude of the laser,
and $\pm \phi$, assumed small, are the phase shifts imposed by the
electro-optic modulators. The operator $\nu(t)$
represents vacuum fluctuations as in Sec.~\ref{secils}.
The second function of the interferometer is to produce a local
oscillator beam with mean amplitude
\beq
-(1/2)(\beta e^{i\phi} + \beta e^{-i\phi}) = -\beta[ 1 +O(\phi^{2})].
\eeq
With appropriate phase shifts assumed, this local oscillator is then
used for making a homodyne measurement of the $X=b+b\dg$ quadrature
of
$b_{\rm out}$ (which, in the absence of the atom, is identical with
$b_{\rm in}$). If the efficiency of the detectors is $\ve$ then the
homodyne photocurrent is represented by the operator \cite{Gar91}
\beq \label{homI}
I_{\rm hom}(t) = \rt{\ve}X_{\rm out}(t)
+ \rt{1-\ve}\xi_{\ve}(t),
\eeq
where $\xi_{\ve}(t)$ is a unit-norm real white noise process.

Although $S^{X}_{\rm out}(\omega)$ is
shot-noise limited for high frequencies, it need not be for lower
frequencies. In particular, the feedback loop shown can produce a
spectrum below  the shot-noise as shown in Sec.~\ref{secils}.
The current $I_{\rm hom}(t)$
is amplified and used to control $\phi$. If we set
\beq \label{setphi}
\phi(t) = \frac{g}{2\beta\sqrt{\ve}}\int_{0}^{\tau}h(t') 
I_{\rm hom}(t-T-t') dt',
\eeq
then we have a feedback loop
with a low-frequency round-loop gain of $g$
which is stable under the conditions given in Sec.~\ref{sec:stab}. Solving in the Fourier domain,
\beq
\tilde{X}_{\rm out}(\omega) =
\left[{\tilde{\xi}_{\nu}(\omega) +
g\tilde{h}(\omega)e^{-i\omega
T}\rt{\frac{1-\ve}{\ve}}\tilde{\xi_{\ve}}(\omega)}\right]
\frac{1}{1-g\tilde{h}(\omega)e^{-i\omega T}}.
\eeq
Thus $X_{\rm in}$ (which here equals $X_{\rm out}$) has a spectrum
\beq \label{spec}
S^{X}_{\rm in}(\omega) =
[{1+g^{2}|\tilde{h}(\omega)|^{2}(\ve^{-1}-1)}]/{|1-g\tilde{h}
(\omega)e^{-i\omega T}|^{2}}.
\eeq
At a frequency $\bar\omega$ much less than the feedback bandwidth
$\sim \tau^{-1}$, and much less than the reciprocal of the delay 
time $T^{-1}$, $\tilde{h}(\bar\omega)e^{-i\omega T} \simeq 1$ 
and the minimum noise is
\beq \label{minin}
S^{X}_{\rm in}(\bar\omega)_{\rm min} = 1-\ve \;,\;\; {\rm for}\;
g= - \ve/(1-\ve),
\eeq
which is clearly below the standard quantum limit. These results
reproduces
those of \erf{Sminfor1}, where here $\ve$ is playing the same role
as $\eta_{2}$.

\subsection{An In-Loop Atom}

Returning to Fig.~\ref{wfig3}, we now include the two-level atom,
which is assumed to be resonant to the laser. It couples strongly
only to modes of the radiation field having the appropriate dipole
spatial distribution \cite{WalMil94}.
However, by focusing a beam as shown in Fig.~\ref{wfig3}, it is
possible to mode-match a significant proportion, say $\eta$, of
$b_{\rm in}$ into the atom's input.  Recent numerical calculations 
indicate that practical schemes for focusing light in free space 
have a limit on $\eta$ of 
order $0.1$ \cite{EnkKim99}. This suggests that in practice, a more efficient
way to increase the effective $\eta$ would be to couple the light into a
microcavity, as in~\cite{Ver98}. However, it is conceptually simpler 
to consider the free-space set up in Fig.~\ref{wfig3}. 

The Hamiltonian of the atom in
the interaction picture at time $t$ is 
\beq  \label{Hfun}
H(t) = -i[\rt\eta b_{\rm in}(t) +
\rt{1-\eta}\mu(t)]\sigma\dg(t) + {\rm H.c.}
\eeq
Here $\sigma = \ket{g}\bra{e}$ is the atomic lowering operator and
I have set the longitudinal atomic decay rate to unity. The
operator $\mu(t)$ represents an independent vacuum input. Under this
coupling, the output field is found from the techniques of
Sec.~\ref{secqle} to be
\beq \label{ior}
b_{\rm out}(t) = b_{\rm in}(t) + \rt\eta \sigma(t).
\eeq

Although it would be possible to give a description of the entire
feedback loop in terms of atomic and radiation field operators 
\cite{Wis98b}, it is
simpler to use the quantum trajectory theory of homodyne measurement
as outlined in Sec.~\ref{secqt}. In this theory, only the atom is
treated as
a quantum mechanical system with state matrix $\rho(t)$;
the rest of the apparatus is considered
as a complicated measurement and feedback device for the atom. The
photocurrent $I_{\rm c}^{\rm hom}(t)$ is therefore a classical quantity. From
\erf{422hompc2} it is given by
\beq \label{ce1}
I_{\rm c}^{\rm hom}(t) = \bar{I}_{\rm c}^{\rm hom}(t) + \xi^{\rm hom}(t),
\eeq
where $\xi^{\rm hom}(t)$ is local-oscillator shot noise, which in
this case is the only source of noise in the whole system.
From~(\ref{homI}) and (\ref{ior}), the expected value
$\bar{I}_{\rm c}^{\rm hom}(t)$ conditioned upon the prior 
photocurrent record is
\beq \label{ce2}
\bar{I}_{\rm c}^{\rm hom}(t) = \rt{\eta\ve} {\rm Tr}[\rho_{\rm c}(t)\sigma_{x}] +
\rt{\ve}2\beta\phi_{\rm c}(t).
\eeq
Here $\phi_{\rm c}(t)$ is not set to its average value of zero because
it is 
determined by the prior classical photocurrent via \erf{setphi}.

From the theory of Sec.~\ref{secqt}, the atom will obey the
following nonlinear stochastic master equation
\beq \label{912}
d\rho _{\rm c} = dt{\cal D}[\sigma]\rho_{\rm c}
 + \rt{\eta\ve}dW^{\rm hom}(t){\cal H}[\sigma]\rho_{\rm c}
-idt[H_{\rm fb},\rho_{\rm c}].
\eeq
The final Hamiltonian is due to the feedback. It is
identical to the term due to feedback in the fundamental atomic
Hamiltonian \erf{Hfun}, namely
\beq \label{sub1}
H_{\rm fb}(t) =  \rt{\eta} \beta\phi(t) \sigma_{y}.
\eeq
So far, it is not a 
singular quantity because of the smoothing of the 
photocurrent in \erf{setphi}.

Now consider the limit of instantaneous feedback on the atomic
time-scale, $\tau, T \ll 1$. In this limit
\beq \label{sub2}
2\beta\phi(t) = gI_{\rm c}^{\rm hom}(t)/\sqrt{\ve}
\eeq
and thus we can derive
\beq
I_{\rm c}^{\rm hom}(t) = (1-g)^{-1}\{\xi(t) + \rt{\eta\ve}
{\rm Tr}[\rho_{\rm c}(t)\sigma_{x}]\}.
\eeq
Hence from~(\ref{sub1}) and (\ref{sub2}),
\beq \label{916}
H_{\rm fb}(t) = \lambda \half \sigma_y \{
 {\rm Tr}[\rho_{\rm c}(t)\sigma_{x}]+\xi(t)/\rt{\eta\ve}\},
\eeq
where it is to be understood that $t$ on the right-hand side of
this equation actually stands for $t-0^+$. Since $-\infty < g < 1$, 
the feedback parameter
$\lambda$ is given by
\beq
\lambda = {g\eta}/({1-g}) \; \in (-\eta,\infty).
\eeq

Taking the Markovian limit allows us to derive a deterministic master
equation for the atom. Equation (\ref{912}) no longer applies, 
because $H_{\rm fb}(t)$ in \erf{916} is singular. 
However, by comparison of \erf{916} with \erf{Hfb} we see that 
 we can follow the methods of Sec.~\ref{sec:hfb} 
and Sec~\ref{secfme}
to obtain 
\beq \label{cenme}
\dot\rho = {\cal D}[\sigma]\rho
-i\lambda[\half\sigma_y,\sigma \rho + \rho\sigma\dg]
+ \frac{\lambda^2}{\eta\ve}{\cal D}[\half\sigma_y]\rho.
\eeq
This equation, and
the following relation between $\lambda$ and the in-loop squeezing
\beq \label{sbw}
S^{X}_{\rm in}(\bar\omega) =
\frac{1+g^{2}(\ve^{-1}-1)}{(1-g)^{2}}
= 1+\frac{2\lambda}{\eta}+\frac{\lambda^{2}}{\eta^{2}\ve},
\eeq
are the central results of this section. In \erf{sbw}, we still have
$\bar\omega \ll \tau^{-1}$, but now also
 $\bar\omega\gg 1$. This ensures that the atomic variables
(with characteristic time scale of unity) do not contribute
significantly to the
spectrum at $\bar\omega$, so that \erf{spec} is still valid.

From the master equation \erf{cenme} it is easy to derive the
following dynamical equations:
\bqa
{\rm Tr}[\dot\rho \sigma_{x}] &=& -\gamma_{x}{\rm Tr}[\rho
\sigma_{x}] \label{onex}\\
{\rm Tr}[\dot\rho \sigma_{y}] &=& -\gamma_{y}{\rm Tr}[\rho
\sigma_{y}]\\
{\rm Tr}[\dot\rho \sigma_{z}] &=& -\gamma_{z}{\rm Tr}[\rho
\sigma_{z}]-C \label{threez}
\eqa
Only the equation for $\sigma_{y}$ is unaffected by the feedback,
with $\gamma_{y} = 1/2.$  The new decay rate for $\sigma_{x}$ is
\beq
\gamma_{x} = \frac12\left[ 1+2\lambda +
\frac{\lambda^{2}}{\eta\ve}\right],
\eeq
and the modified parameters for $\sigma_{z}$ are
\beq
\gamma_{z} = \gamma_{y} + \gamma_{x}\;,\;\;C = 1+\lambda.
\eeq
In steady state
\bqa
{\rm Tr}[\rho_{\rm ss}\sigma_{x}]&=&
{\rm Tr}[\rho_{\rm ss}\sigma_{y}]=0 , \\
{\rm Tr}[\rho_{\rm ss}\sigma_{z}] &=& -1 +
{\lambda^{2}}/[{2\eta\ve(1+\lambda)+\lambda^{2}}].
\eqa

The most interesting of these results is that
negative feedback can reduce
the decay rate of the $x$ component of the atomic dipole
below its natural value of $1/2$.
From \erf{sbw} it can be re-expressed as
\beq \label{gx1}
\gamma_{x} = \half \left[ (1 - \eta) + \eta S^{X}_{\rm
in}(\bar\omega)\right] .
\eeq
This clearly shows that $\gamma_{x}$ has two contributions:
$\half(1-\eta)$ from the vacuum input and $\half \eta S^{X}_{\rm
in}(\bar{\omega})$ from the in-loop squeezed light.
The greatest
reduction occurs for minimum in-loop fluctuations
as in \erf{minin}, for which
\beq
(\gamma_{x})_{\rm min} = \half\left( 1- \eta\ve \right) \;,\;\;
{\rm for}\;\lambda = -\eta\ve.
\eeq

\begin{figure}
\includegraphics[width=1\textwidth]{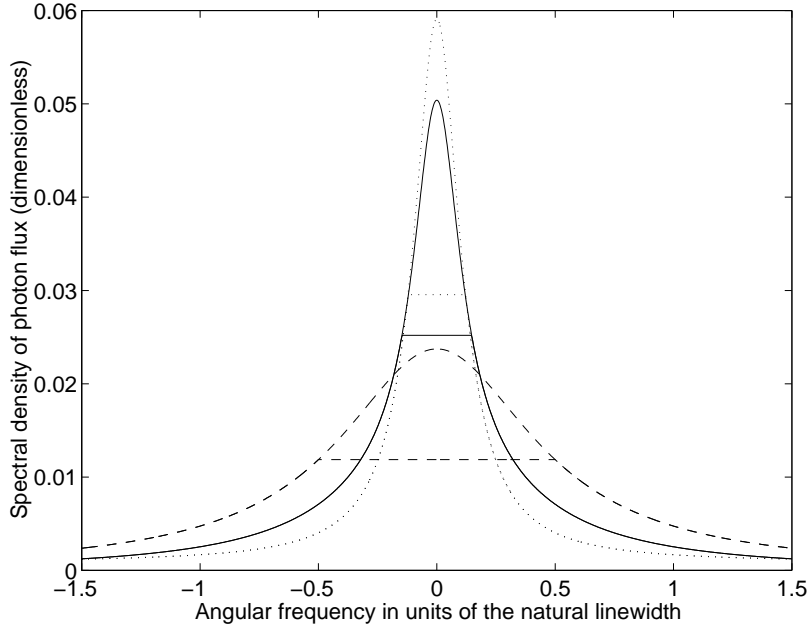}
	\caption{Plot of the Power Spectrum $P(\omega)$ of the
	fluorescence into the vacuum modes, for  in-loop squeezing
	(solid) and free squeezing (dotted), with
	mode-matching $\eta=0.8$ and squeezing
	$S^{X}_{\rm in}(\bar\omega)=0.05$. The linewidth for
	in-loop squeezing is slightly broader
	because the contribution from $\sigma_{y}$
	is {\em not} broadened in this case. The
	natural-width spectrum from weak thermal driving (dashed)
	is scaled up for comparison}
	\protect\label{wfig4}
\end{figure}

The slower decay of $\sigma_{x}$ can be directly observed in the
power spectrum of the fluorescence of the atom into the vacuum modes.
This measures the photon flux per unit frequency into these modes
and is defined by
\beq
P(\omega) = \frac{1-\eta}{2\pi}\ip{\tilde{\sigma}\dg(-\omega)
\sigma(0)}_{\rm ss}.
\eeq
From \erf{onex}--\erf{threez} this is easily evaluated to be
\beq \label{genps}
P(\omega) = \frac{(1-\eta)(\gamma_{z}-C)}{8\pi\gamma_{z}}\left[
\frac{\gamma_{x}}{\gamma_{x}^{2}+\omega^{2}} +
\frac{\gamma_{y}}{\gamma_{y}^{2}+\omega^{2}}\right].
\eeq
For the optimal squeezing ($\lambda =-\eta\ve$) we have
\beq
P(\omega) = \frac{(1-\eta)\eta\ve}{4\pi(2 - \eta\ve)} \left[
\frac{1-\eta\ve}{(1-\eta\ve)^{2}+4\omega^{2}} +
\frac{1}{1+4\omega^{2}}\right].
\eeq
This is plotted in Fig.~\ref{wfig4} for $\eta=0.8$ and $\ve=0.95$.

\subsection{Comparison with Free Squeezing.}

To compare the above results with those produced by free squeezing we
again assume that the mode-matching of the squeezed modes into the
atom is
$\eta$, and that the squeezing is broad-band compared to the atom.
Assuming also that the input light is in a
minimum-uncertainty state for the $X$ and $Y$ quadratures
\cite{WalMil94},
it can be characterized by a single real number $L$, with
\beq\label{L}
S_{\rm in}^{X}(\omega) = L = 1/S_{\rm in}^{Y}(\omega) .
\eeq
This parameter is related to those of Sec.~\ref{TomVit} by
 $L=2N+2M+1$, where $M^{2}=N(N+1)$ and $M$ is assumed real.
This yields the master equation
\beq
\dot\rho = (1-\eta){\cal D}[\sigma]\rho + \frac{\eta}{4L}
{\cal D}[(L+1)\sigma-(L-1)\sigma\dg]\rho,
\eeq
which leads again to (\ref{onex})--(\ref{threez}), but with
\bqa
\gamma_{x} &=& \half\left[ (1-\eta)+\eta L\right] \label{gx2},\\
\gamma_{y} &=& \half\left[ (1-\eta)+\eta L^{-1}\right], \\
\gamma_{z} &=& \gamma_{x}+\gamma_{y} \;,\;\;C=1.
\eqa

For $L<1$ the decay of  $\sigma_{x}$ is again inhibited.
The crucial observation to be made is that the dependence of
$\gamma_{x}$ on the degree of $X$ quadrature squeezing of the input
light is exactly the same as for in-loop squeezing, as is seen by
comparing~(\ref{L}) and (\ref{gx2}) with \erf{gx1}. The only
difference between the two cases is that $C$ is unaffected by the
free squeezing and that $\gamma_{y}$ is not increased by the in-loop
squeezing. The latter is a direct consequence of the fact that an
in-loop field is not bound by the usual two-time uncertainty
relations. The
free squeezing fluorescence spectrum is again given by \erf{genps}.
This is also plotted in Fig.~\ref{wfig4} for $\eta=0.8$ and $L=0.05$.
As this figure shows, the spectra are certainly not identical,
but the sub-natural linewidth is much the same in both.

To conclude, line-narrowing of an
atom is not a diagnostic of free squeezing.  Rather, it requires
only temporal anticorrelations of one
quadrature of the input field (for times much shorter than the
atomic lifetime) such as can be produced by a negative electro-optic
feedback loop.  The dependence of the line-narrowing on the input
squeezing and the degree of mode-matching is the same for in-loop
squeezing as for free squeezing.  Because the quadrature operators of
an in-loop field do not obey the usual two-time commutation
relations,
the reduction in noise in one quadrature does not imply an increase
in noise in the other.  Hence the line-narrowing of one quadrature of
the atomic dipole by in-loop squeezing does not entail the
line-broadening
of the other quadrature. What significance this difference has in
the physics of more complex atomic interactions with squeezed light
\cite{FicDru97} is a question requiring much investigation.

In-loop squeezing is generally easier to produce than free squeezing
for a number of reasons.  First, in-loop squeezing does not require
expensive and delicate sources such as nonlinear crystals, but rather
off-the-shelf electronic and electro-optical equipment.  Second, the
amount of squeezing is limited only by the efficiency of the
photodetection.  For homodyne detection, as required here, an
efficiency of 95\% is readily obtainable \cite{Sch96} and would
enable in-loop
squeezing of 95\%.  Third, in-loop squeezing can be produced
at any frequency for which a
coherent source is available, so experiments could be conducted on
any
atomic transition.  The one difficulty with in-loop squeezing is that
it requires a feedback loop response time much shorter than an atomic
lifetime, but this would not be a problem for metastable
transitions.
Thus as well as  giving us a better theoretical understanding
of the effects on matter of light with
fluctuations below the standard quantum limit, in-loop squeezing
should be a practical alternative to free
squeezing in the experimental investigation of these effects.

\subsection{Other Uses of ``Squashed light''}

The analysis given above is probably the most interesting application
for squashed light proposed so far. However, it is worth pointing out
that another conceivable application was suggested in~\cite{Tau95}.
This was in the context of a QND measurement of an the squashed
in-loop quadrature. For a perfect QND device, such a measurement does
not disturb the feedback loop. Now because the in-loop fluctuations
are squashed, the readout from the QND device will have less noise
than it would have without feedback. This is useful if the
measurement
is being made with the aim not of determining the in-loop intensity,
but of determining the coupling constant between the in-loop light
and the QND meter. For a given detector bandwidth, the squashing of
intensity fluctuations would thus enable this parameter to be
estimated more accurately. The improvement in accuracy would be
limited only by the efficiency of the photodetectors used in the
feedback loop, as in \erf{minin}.

Another potential application for squashed light was proposed
recently in~\cite{Buc99}. Here the idea is to reduce radiation
pressure fluctuations on a mirror by squashing the amplitude
fluctuations. The feedback involves standard photodetection and an
amplitude modulator. The point of reducing the radiation pressure
fluctuations is again to allow a more accurate estimation of other
parameters, namely the spectrum of the thermal noise in the mirror.
In this case it turns out that in the optimal regime the fluctuations
can at most be reduced by a factor of one half.

\section{Conclusion}

What then, can we say in conclusion about squeezing and feedback? The
first and foremost fact is that, in the absence of any nonlinear
optical elements, feedback cannot produce free squeezing. A nonlinear
optical element is any element whose effect cannot be modeled by a
displacement, rotation, or damping of the amplitude of the light.
By free squeezing I mean squeezing whose existence can be verified by
detection in a conventional (demolition) detector which is not part
of the feedback loop. This includes both continuum squeezing and
intracavity squeezing (which can be measured by dumping the light out
of the cavity).

If there are nonlinear optical elements present then there are many
interesting things one can achieve with feedback involving squeezing.
First, if those elements are used to generate a
squeezed beam which is split, then feedback (or indeed feedforward)
onto an intensity modulator
can effectively transfer amplitude squeezing electronically from the
beam which is detected to the free beam. This was discussed in
Sec.~\ref{secasi}.
Second, nonlinear optical elements can create quantum correlations
between two beams. Then, even if the beams separately are not
squeezed, feedback (or, equally effectively, feedforward)
of the light from one beam can make the other beam squeezed. If they
are squeezed, then feedback or feedforward can enhance this squeezing.
A number of scenarios were discussed in
Sec.~\ref{secfnm}. Third, if the nonlinear element creates squeezing
within a cavity then controlling the driving of the cavity by a
photocurrent derived from a homodyne measurement of the output
can enhance the squeezing, as discussed in Sec.~\ref{secics}.
This could be seen as a generalization of the first case discussed
above, where the cavity mirror is acting as a beam splitter and the
intracavity light is being continually split.

A quite separate issue is the nature of the light within a feedback
loop.  At first sight it appears easy to make this squeezed,
 as negative feedback (without the aid of any optical nonlinearity)
can produce an {\em in-loop} photocurrent with arbitrarily low noise,
regardless of the efficiency of the detector.
However, the fact that this is the current from an in-loop detector
means that a sub-shot-noise spectrum does not have the usual
significance. First, there is a perfectly good semiclassical
explanation for this phenomenon in terms of coherent field states.
Second, the usual two-time commutation relations for a free field do
not hold for an in-loop field. This means that there is no
``squeezing'' of the uncertainty from the amplitude  to the phase
quadrature, rather
just an apparent ``squashing'' of the amplitude uncertainty with no
effect on the phase uncertainty. Lastly, from a practical point of
view, the squeezed light cannot be removed from the loop using a
linear optical device. A beam splitter inserted in the in-loop beam
produces a free (out-of-loop)
beam with noise level above rather than below the shot noise.

Despite the differences between in-loop squashing and free squeezing,
it turns out that there are similarities. An in-loop quantum
non-demolition amplitude detector will respond to amplitude squashing
exactly as to amplitude squeezing. For a feedback loop constructed 
using a perfectly efficient
(demolition) detection, a perfect in-loop non-demolition
detector will have a read-out identical to that of the in-loop
demolition detector, and, for arbitrarily large negative feedback,
this will have an arbitrarily low noise level. For non-unit efficiency
of the demolition detector in the feedback
loop, the two detectors will not agree and the maximum degree of
squashing observed by the non-demolition detector will be limited by the
demolition detector efficiency.

The degree of squashing seen by a perfect QND detector turns out to
be a legitimate measure of the degree of noise reduction in the light,
and this squashing can have effects on nonlinear optical devices
very similar to those produced by squeezing. In particular, shining
broad-band quadrature squashed light on a resonant two-level atom will
cause the decay of  the in-phase quadrature of the atomic dipole
to be suppressed. This manifest itself as a line-narrowing in the
power spectrum of the atom's fluorescence, an effect which was originally
thought to be characteristic of squeezing.  Moreover, the reduction
in the linewidth of the in-phase quadrature depends on the degree of
squashing (as measured by a QND device) in precisely the same way as
it does on the degree of squeezing in the case of freely propagating
light. This has important experimental implications as there are are
a number of factors which make squashing easier to achieve than
squeezing.

Very recently, there have been more intriguing developments in the
theory of squashed light and its application to quantum spectroscopy
\cite{Wis99}. First, I showed that it is possible to squash light
simultaneously in both quadratures, and also to squeeze light in one
quadrature and squash it the other. In the limit of perfect squeezing,
and perfect squashing (which requires unit efficiency detectors) the
in-loop fluctuations can be banished from both quadratures! An atom
coupled only to this in-loop light would (to a first approximation)
have its spontaneous decay completely suppressed. This surprising
prediction is one more example of the continuing
fruitful investigation of the relationship between feedback and squeezing.

\subsection*{Acknowledgment}
I would like to thank Laura Thomsen for a careful reading of this 
manuscript. In work subsequent to the completion of this manuscript, we 
have generalized the work described above by considering the effect of twin-beam squashed light 
on a three-level atom \cite{ThoWis01}. We have also shown how feedback can be used to prepare near-minimum uncertainty spin-squeezed states \cite{ThoManWis02}.


\begin{thebibliography}{7}

\bibitem{WalMil94}
D. F. Walls and G. J. Milburn,
{\em Quantum Optics}
(Springer, Berlin, 1994).

\bibitem{Gar91}
C. W. Gardiner,
{\em Quantum Noise}
(Springer, Berlin, 1991).

\bibitem{WisMil94b} 
H. M. Wiseman and G. J. Milburn,
Phys. Rev. A {\bf 49}, 4110 (1994).

\bibitem{WalJak85a}
J. G.~Walker and E.~Jakeman,
Proc. Soc. Photo-Opt. Instrum. Eng. {\bf 492}, 274 (1985).

\bibitem{MacYam85}
S.~Machida and Y.~Yamamoto,
Opt. Commun. {\bf 55}, 219 (1985).

\bibitem{HauYam86}
H. A. Haus and Y. Yamamoto,
Phys. Rev. A {\bf 34}, 270 (1986).

\bibitem{YamImoMac86}
Y. Yamamoto, N. Imoto and S. Machida,
Phys. Rev. A {\bf 33}, 3243 (1986).

\bibitem{Sha87}
J. M. Shapiro {\em et al},
J. Opt. Soc. Am. B {\bf 4}, 1604 (1987).

\bibitem{CarTiaRenAls94}
H. J. Carmichael, L. Tian, W. Ren and P. Alsing,
in {\em Cavity QED}, ed. Paul Berman, vol. 34 of
Advances in AMO Physics (1994).

\bibitem{WisMil93b}
H. M. Wiseman and G. J. Milburn,
Phys. Rev. Lett. {\bf 70}, 548 (1993).

\bibitem{WisMil94a} 
H. M. Wiseman and G. J. Milburn,
Phys. Rev. A {\bf 49}, 1350 (1994).

\bibitem{Wis94a} 
H. M. Wiseman,
Phys. Rev. A {\bf 49}, 2133 (1994).

\bibitem{Car93b}
H. J. Carmichael,
{\em An Open Systems Approach to Quantum Optics}
(Springer, Berlin, 1993).

\bibitem{GarParZol92}
C. W. Gardiner, A. S. Parkins and P. Zoller,
Phys. Rev. A {\bf 46}, 4363 (1992).

\bibitem{WisMil93c} 
H. M. Wiseman and G. J. Milburn,
Phys. Rev. A {\bf 47}, 1652  (1993).

\bibitem{Pli94}
L. Plimak,
Phys. Rev. A {\bf 50}, 2120 (1994).

\bibitem{Gla63a}
R. J. Glauber,
Phys. Rev. {\bf 130}, 2529 (1963).

\bibitem{Gla63b}
R. J. Glauber,
Phys. Rev. {\bf 131}, 2766 (1963).

\bibitem{Sud63}
E. C. G. Sudarshan,
Phys. Rev. Lett. {\bf 10}, 277 (1963).

\bibitem{Tro91}
A. S. Troshin,
Opt. Spektrosc. (USSR) {\bf 70}, 389 (1991).

\bibitem{HeiMer93}
A. Heidmann and J. Mertz,
J. Opt. Soc. Am. B {\bf 10}, 1637 (1993).

\bibitem{CohDupGry89}
C. Cohen-Tannoudji, J. Dupont Roc and G. Grynberg,
{\em Photons and Atoms: An Introduction to Quantum Electrodynamics}
(Wiley, New York, 1989).

\bibitem{GarCol85}
C. W. Gardiner and M. J. Collett,
Phys. Rev. A {\bf 31}, 3761 (1985).

\bibitem{Mol96}
K.  M\o lmer,
Phys. Rev. A {\bf 55}, 3195 (1996).

\bibitem{YueSha80}
H. P. Yuen and J. H. Shapiro,
IEEE Trans. IT {\bf 26}, 78 (1980).

\bibitem{Tau95}
M.~S.~Taubman {\em et al.}
J. Opt. Soc. Am. B {\bf 12}, 1792 (1995).

\bibitem{Mas94}
A.~V.~Masalov, A.~A.~Putilin and M.~V.~Vasilyev,
J. Mod. Opt. {\bf 41}, 1941 (1994).

\bibitem{Ste90}
J. J. Stefano, A. R. Subberud, and I. J. Williams,
{\em Theory and Problems of Feedback and Control Systems 2e}
(Mc Graw-Hill, New York, 1990).

\bibitem{Fon91}
Ya. A. Fananov,
Opt. Spektrosc. (USSR) {\bf 70}, 392 (1991).

\bibitem{You94}
S.-H.~Youn {\em et al.}
J. Opt. Soc. Am. B {\bf 11}, 102 (1994).

\bibitem{Buc99}
B. C. Buchler {\em et al.},
Opt. Lett. {\bf 24}, 259 (1999).

\bibitem{KhoKil94}
D. B. Khoroshko and S. Ya. Kilin,
JETP {\bf 79}, 691 (1994).

\bibitem{Tau96}
M.S. Taubman {\em et al.},
Proceedings of the 12th International Congress on Lasers in
Research and Engineering (Springer, Berlin, 1996).

\bibitem{HilScu82}
M. Hillery and M.O. Scully,
Phys. Rev. D {\bf 25}, 3137 (1982).

\bibitem{Yur85}
B. Yurke,
J. Opt. Soc. Am. {\bf B2}, 732 (1985).

\bibitem{AlsMilWal88}
P. Alsing, G. J. Milburn and D. F. Walls
Phys. Rev. A {\bf 37}, 2970 (1988).

\bibitem{JakWal85}
E.~Jakeman and J. G.~Walker,
Opt. Commun. {\bf 55}, 219 (1985).

\bibitem{WalJak85b}
J. G.~Walker and E.~Jakeman,
Optica-Acta. {\bf 32}, 1303 (1985) 

\bibitem{TapRarSat88} 
P. R. Tapster, J. G. Rarity and J. S. Satchell,
Phys. Rev. A {\bf 37}, 2963 (1988).

\bibitem{Mer90} 
J. Mertz, A. Heidmann, C. Fabre, E. Giacobino and S. Reynaud,
Phys. Rev. Lett. {\bf 64}, 2897 (1990).

\bibitem{MerHeiFab91}
J. Mertz, A. Heidmann and C. Fabre,
Phys. Rev. A {\bf 44}, 3329 (1991).

\bibitem{ColLev91} 
M. J. Collett and R. B. Levien,
Phys. Rev. A {\bf 43}, 5068 (1991)

\bibitem{WisTauBac95}
H. M. Wiseman, M. S. Taubman and H.-A. Bachor,
Phys. Rev. A {\bf 51} 3227 (1995).

\bibitem{Pas94}
R. Paschotta {\em et al},
Phys. Rev. Lett. {\bf 72}, 3807 (1994)

\bibitem{Lin76}
G. Lindblad,
Commun. Math. Phys. {\bf 48}, 199 (1976).

\bibitem{Von32}
J. von Neumann, {\em Mathematical Foundations of Quantum Mechanics}
(Springer, Berlin, 1932);
English translation (Princeton University Press, Princeton, 1955).

\bibitem{DalCasMol92}
J. Dalibard, Y. Castin and K. M\o lmer,
Phys. Rev. Lett. {\bf 68}, 580 (1992).

\bibitem{MolCasDal93}
K. M\o lmer, Y. Castin and J. Dalibard,
J. Opt. Soc. Am. B {\bf 10}, 524 (1993).

\bibitem{DumZolRit92}
R. Dum, P. Zoller and H. Ritsch,
Phys. Rev. A {\bf 45}, 4879 (1992).

\bibitem{DumParZolGar92}
R. Dum, A. S. Parkins, P. Zoller and C. W. Gardiner,
Phys. Rev. A  {\bf 46}, 4382 (1992).

\bibitem{TiaCar92} 
L. Tian and H. J. Carmichael,
Phys. Rev. A {\bf 46}, R6801 (1992).

\bibitem{MarDumTaiZol93}
P. Marte, R. Dum, R. Ta\"{\i}eb and P. Zoller,
Phys. Rev. A {\bf 47}, 1378 (1993).

\bibitem{WisMil93a} 
H. M. Wiseman and G. J. Milburn,
Phys. Rev. A {\bf 47}, 642 (1993).

\bibitem{Gar85}
C. W. Gardiner,
{\em Handbook of Stochastic Methods}
(Springer, Berlin, 1985).

\bibitem{GioTomVit99}
V. Giovannetti, P. Tombesi and D. Vitali,
quant-ph/9902077

\bibitem{TomVit94}
P. Tombesi and D. Vitali,
Phys. Rev. A {\bf 50}, 4253 (1994).

\bibitem{FilJavMey86}
P. Filipowicz, J. Javanainen and P. Meystre,
Phys. Rev. A {\bf 34}, 3077 (1986).

\bibitem{LieMil95}
A. Liebman and G. J. Milburn,
Phys. Rev. A {\bf 51}, 736 (1995).

\bibitem{Gar86}
C. W.~Gardiner,
Phys. Rev. Lett. {\bf 56}, 1917 (1986).

\bibitem{FicDru97}
Z. Ficek and P. D. Drummond,
Phys. Today {\bf 50}, 34 (1997).

\bibitem{Wis98b}
H. M. Wiseman,
Phys. Rev. Lett. {\bf 81}, 3840 (1998).

\bibitem{EnkKim99}
S.J. van Enk and H.J. Kimble,
quant-ph/9908082.

\bibitem{Ver98}
D. W. Vernoy {\em et al.},
Phys. Rev. A {\bf 57}, R2293 (1998).

\bibitem{Sch96}
S. Schiller {\em et al.},
Phys. Rev. Lett. {\bf 77}, 2933 (1996).

\bibitem{Wis99}
H. M. Wiseman,
J. Opt. B: Quant. Semiclass. Opt. {\bf 1}, 459 (1999).

\bibitem{ThoWis01}
L.K. Thomsen and H.M. Wiseman,
	Phys. Rev. A {\bf 64}, 043805 (2001).

\bibitem{ThoManWis02}
L.K. Thomsen, S. Mancini, and H.M. Wiseman,
	Phys. Rev. A (Rapid Comm.) {\bf 65}, 061801 (2002).

\end{thebibliography}
\end{document}